\documentclass[namedreferences]{kluwer}    
\usepackage{graphicx}
\usepackage{epsfig}
\newdisplay{guess}{Conjecture}

\begin{document}

\newcommand{\mic}{$\mu$m}
\newcommand{\hii}{H{\sc ii}}
\newcommand{\hi}{H{\sc i}}
\newcommand{\hal}{H$\alpha$}
\newcommand{\cii}{[C{\sc ii}]}

\begin{article}
\begin{opening}
\title{Normal Nearby Galaxies\thanks{Based
          on observations with ISO, an ESA project
          with instruments funded by ESA Member States (especially
          the PI countries: France, Germany, the Netherlands
          and the United Kingdom) and with the participation
          of ISAS and NASA.}}
\author{Marc \surname{Sauvage}$^{1}$, Richard J. \surname{Tuffs}$^{2}$,
Cristina  C. \surname{Popescu}$^{2}$}
\runningauthor{Sauvage, Tuffs \& Popescu}
\runningtitle{Normal Nearby Galaxies}
\institute{$^{1}$CEA/DSM/DAPNIA/Service d'Astrophysique, \\
C.E. Saclay, 91191 Gif sur Yvette CEDEX, France \\
Email:msauvage@cea.fr\\
$^{2}$Max Planck Institut f\"ur Kernphysik, Astrophysics Department, \\
Saupfercheckweg 1, 69117 Heidelberg, Germany,\\
Email: Richard.Tuffs@mpi-hd.mpg.de\\
 Cristina.Popescu@mpi-hd.mpg.de}

\begin{abstract}
Following on from IRAS, ISO has provided a huge advancement in our
knowledge of the phenomenology of the infrared (IR) emission of
normal galaxies and the underlying physical processes. Highlights
include: the discovery of an extended cold dust emission
component, present in all types of gas-rich galaxies and carrying
the bulk of the dust luminosity; the definitive characterisation
of the spectral energy distribution in the IR, revealing the
channels through which stars power the IR light; the derivation of
realistic geometries for stars and dust from ISO imaging; the
discovery of cold dust associated with HI extending beyond the
optical body of galaxies; the remarkable similarity of the near-IR
(NIR)/ mid-IR (MIR) SEDs for spiral galaxies, revealing the
importance of the photo-dissociation regions in the energy budget
for that wavelength range; the importance of the emission from the
central regions in shaping up the intensity and the colour of the
global MIR luminosity; the discovery of the ``hot'' NIR continuum
emission component of interstellar dust; the predominance of the
diffuse cold neutral medium as the origin for the main
interstellar cooling line, [CII]~158\,\mic,  in normal galaxies.
\end{abstract}

\end{opening}
\noindent{\bf Received: } 30 August 2004, {\bf Accepted: } 29 September 2004

\section{Introduction}

Whereas IRAS provided the first systematic survey of infrared (IR)
emission from normal galaxies, it has been the photometric,
imaging and spectroscopic capabilities of ISO (\opencite{mfk96};
\opencite{mfk03}) which have unravelled the basic physical
processes giving rise to this IR emission. Thanks to the broad
spectral grasp of ISO, the bulk of the emission from dust could be
measured, providing the first quantitative assessment of the
fraction of stellar light re-radiated by dust. The battery of
filters has led to a definitive characterisation of the spectral
energy distribution (SED) in the IR, revealing the contribution of
the different stellar populations in powering the IR emission. The
imaging capabilities have unveiled the complex morphology of
galaxies in the IR, and their changing appearance with IR
wavelength. They also allowed the exploration of hitherto
undetected faint diffuse regions of galaxies. The contribution of
different grain populations to the emission has been measured
through their characteristic spectroscopic signatures. Knowledge
of the emission from cooling lines of the interstellar medium
(ISM) has been extended to low luminosity quiescent spiral and
dwarf galaxies.

In this review we will concentrate on the mid-IR (MIR) to
far-IR (FIR) properties of normal nearby galaxies. By normal
we essentially mean that their SEDs are not powered by accretion.
We will begin with spiral galaxies, since these have attracted most of the
ISO observers' attention. From these objects we
will move to the other class of gas-rich galaxies, the dwarfs. However, as
the nearby extragalactic population is not only made of spirals and dwarfs, we
will conclude by the exploring the very varied IR properties of
early-type galaxies.

\section{Spiral Galaxies}
\subsection{Spatial distributions}
\label{ssec:spatdist}
\subsubsection{FIR Morphologies}
\label{sssec:firmorph}

ISOPHOT (\opencite{dl96}; \opencite{rjl03}) imaged three nearby
galaxies (M~31: Haas et al. 1998; M~33: Hippelein et al. 2003 and
M~101: Tuffs \& Gabriel 2003) in the 60 to 200\,${\mu}$m range,
with sufficient linear resolution to easily distinguish between
the main morphological components in the FIR - nucleus, spiral
arms and underlying disk. The main discovery, made possible by the
unprecedented surface brightness sensitivity longwards of
100\,${\mu}$m, was the existence of large amounts of cold dust
associated both with the spiral arms and with the underlying disk.
This dust was too cold to have been seen by IRAS. Furthermore,
ground-based submillimeter (submm) facilities lacked the surface
brightness sensitivity to map the diffuse component of the cold
dust associated with the underlying disk, though they detected the
component associated with the spiral arms (e.g. Bianchi et al.
2000b).

\begin{figure}[htb] 
\includegraphics[scale=0.4]{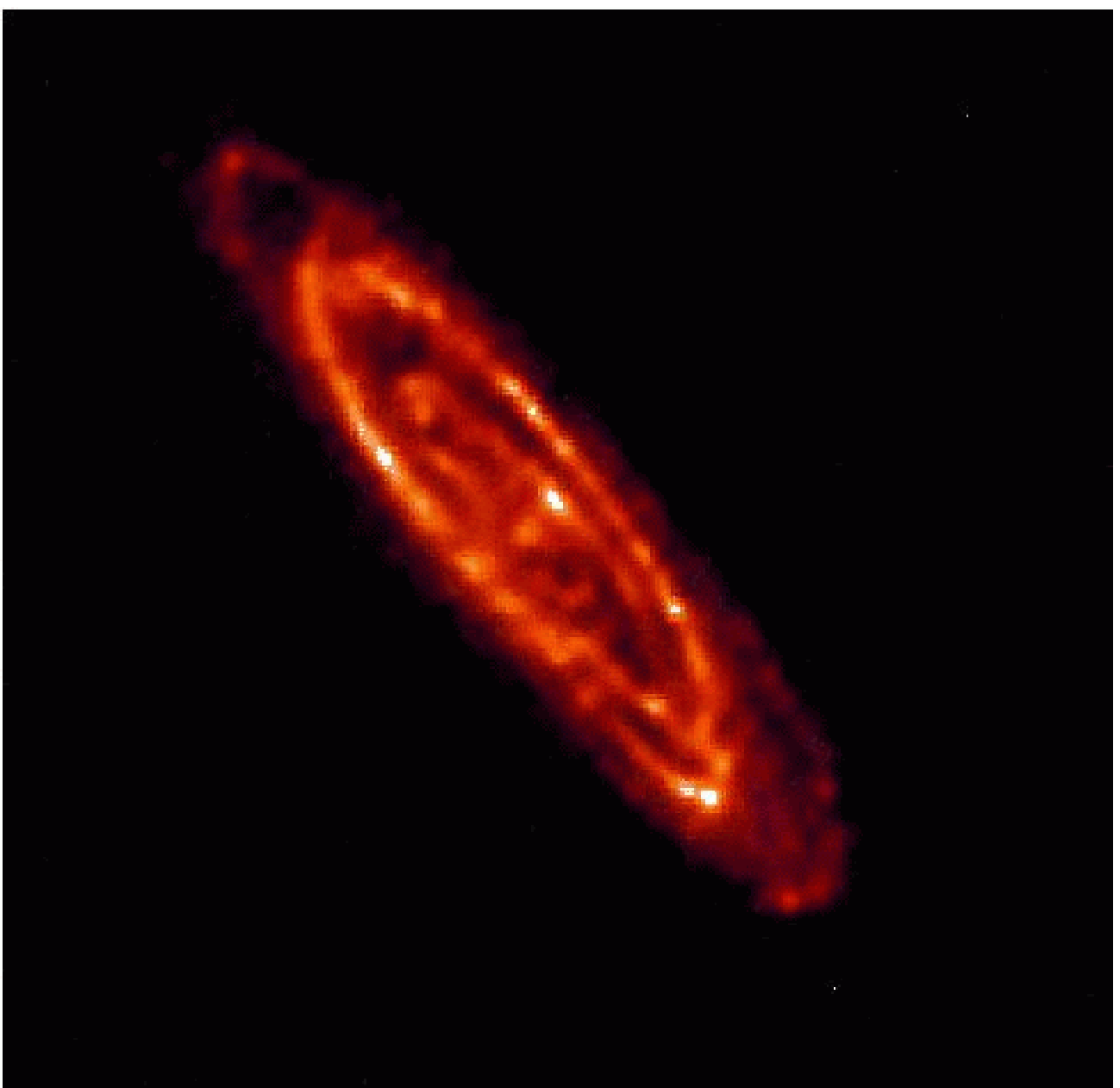}
\includegraphics[scale=0.35]{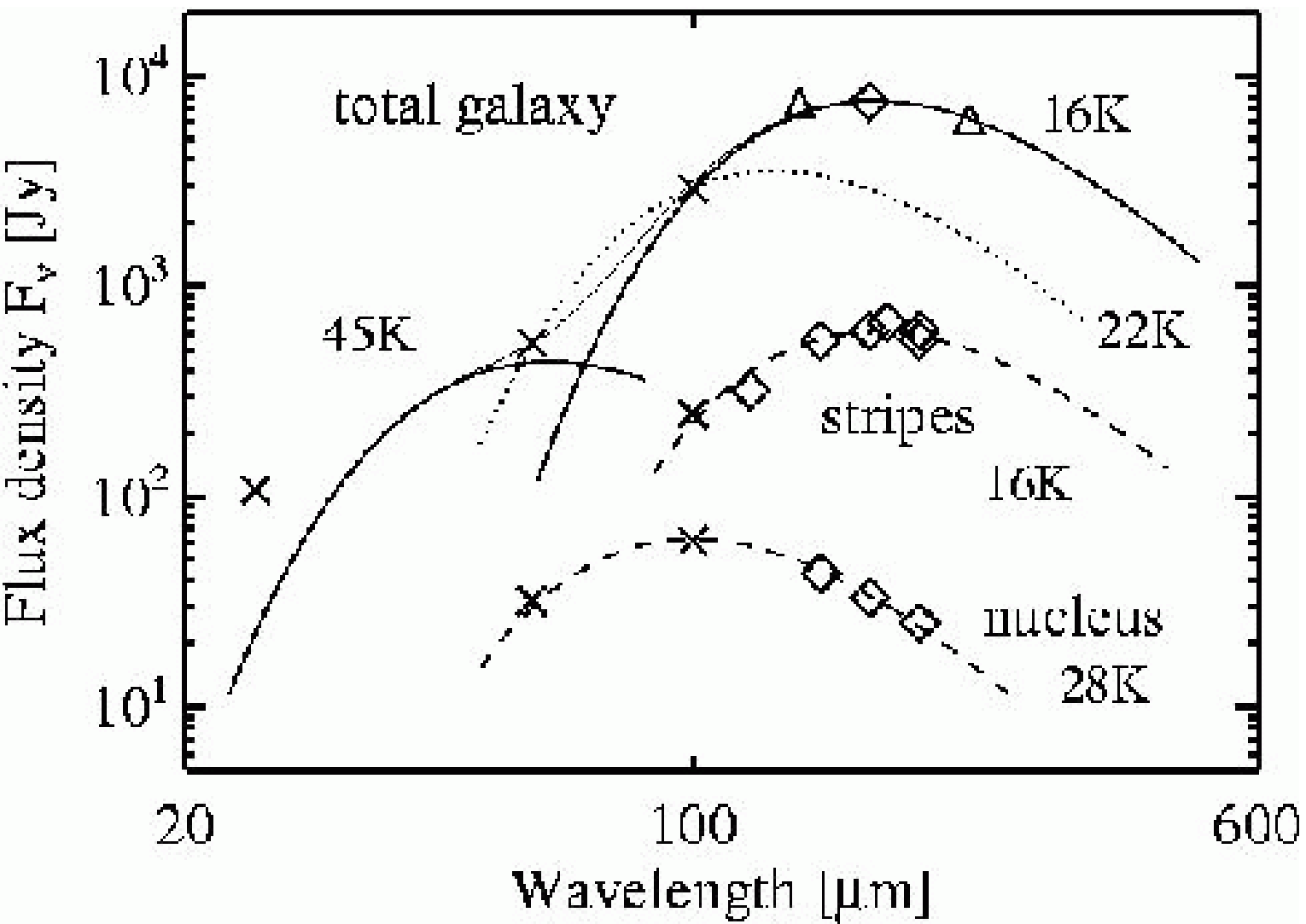}

\caption[]{Left: ISOPHOT 170$\,{\mu}$m map of M~31 (Haas et al. 1998),
with an angular resolution of 1.3$^{\prime}$. North is towards the top,
and East is towards the left. The field size is $2.9 \times 2.9$
degrees. Right: Infrared SED of M~31 (Haas et
al. 1998). The data are shown by symbols (diamonds ISO, crosses IRAS,
triangles DIRBE) with the size being larger than the errors. The
blackbody curves with emissivity proportional to ${\lambda}^{-2}$ are
shown by lines. The dotted line with T=22\,K through the IRAS 60 and
100\,${\mu}$m data points indicates what one would extrapolate from
this wavelength range alone without any further assumptions.}

\label{fig:f1}
\end{figure}

In the case of the Sab galaxy M~31, most of the emission at
170\,${\mu}$m arises from the underlying disk, which has a completely
diffuse appearance (Fig.~\ref{fig:f1}, left panel). This diffuse disk
emission can be traced out to a radius of 22\,kpc, so the galaxy has a
similar overall size in the FIR as seen in the optical bands.
Fig.~\ref{fig:f1} (left panel) also shows that at 170\,${\mu}$m the
spiral arm component is dominated by a ring of 10\,kpc radius.  In
addition, there is a faint nuclear source, which is seen more
prominently in HIRES IRAS 60\,${\mu}$m maps at similar resolution and
in H${\alpha}$.  The overall SED (Fig.~\ref{fig:f1}, right panel)
can be
well described as a superposition of two modified ($\beta$=2) Planck
curves, with dust temperatures $T_{\rm D}$ of 16 and 45\,K.  The cold
dust component at 16\,K arises from both the ring structure (30$\%$)
and the diffuse disk (70$\%$; Haas, private communication),
illustrating the importance of the diffuse emission at least for this
example.  The 45\,K component matches up well with \hii\ regions within
the star-formation complexes in the ring structure. Associated with
each star-formation complex are also compact, cold emission sources
(see Fig.~3 of \opencite{smi00}) with dust
temperatures in the 15 to 20\,K range.  These could well represent the
parent molecular clouds in the star-formation complexes which gave
rise to the \hii\ regions.  Detailed examination of the morphology of
the ring shows a smooth component of cold dust emission as well as the
discrete cold dust sources. Finally, the nuclear emission was fitted
by a 28\,K dust component.

The ISOPHOT maps of the Sc galaxies M~33 and M~101 show the same
morphological components as seen in M~31, with the difference that the
spiral arm structure can be better defined in these later-type
spirals. Also the star-formation complexes in the spiral arms show
similar SEDs to those seen in M~31.

In conclusion, the characteristics of the FIR emission from the main
morphological components of spiral galaxies are:

\begin{itemize}

\item {\bf nucleus}: an unresolved warm source with $T_{\rm
D}\,\sim\,$30\,K

\item {\bf spiral arms}: a superposition of:

\begin{itemize}

\item localised warm emission with 40\,$\le\,T_{\rm D}\,\le\,$60\,K
from \hii\ regions

\item localised cold emission with 15\,$\le\,T_{\rm D}\,\le\,$20\,K
from parent molecular clouds

\item diffuse emission running along the arms

\end{itemize}

\item {\bf disk}: an underlying diffuse (predominantly cold) emission
with 12\,$\le\,T_{\rm D}\,\le\,$20\,K

\end{itemize}

\subsubsection{MIR Morphologies}
\label{sssec:mirmorph}

Generally, ISOCAM (\opencite{cc96}; \opencite{jb03}) maps of
nearby galaxies are difficult to compare with their FIR
counterparts from ISOPHOT, for two reasons. On the one hand
ISOCAM's far superior angular resolution enables the mapping of
detailed structures even within the spiral arm. On the other hand
ISOPHOT's surface brightness sensitivity is far superior to that
of ISOCAM, enabling faint diffuse emission on scales of galactic
disks to be traced. ISOCAM's smaller PSF also meant that different
galaxies were generally mapped in the MIR than in the FIR. M\,31
and M\,33 were too large to be completely mapped with ISOCAM.

Observations of a portion of the southern disk of
M\,31 were presented by \inlinecite{pag99}. Due to the vastly
different scales, these observations are particularly difficult to relate
to those of \inlinecite{haa98} but they nevertheless reveal that the
detected MIR emission originates mostly in the regions of the
spiral arms of Fig.~\ref{fig:f1}. In fact, M\,31 was completely mapped
by \inlinecite{kra02} at a number of MIR wavelengths with the
{\em Midcourse Space Experiment} and their 8.3\,\mic\ map (their
Fig.~1) is strikingly similar to that of \inlinecite{haa98}, with the
difference that the diffuse emission detected in the FIR beyond the
main spiral arms is not detected in the MIR. This difference could for instance
be understood as being due to different parts of the grain size distribution
having different spatial extents. However, generalising such a conclusion on
the basis of the observation of M~31 alone could be dangerous, as it is known
that some of the dust properties of this galaxy are very different from those
of other galaxies, as revealed by it's atypical MIR spectrum \cite{ces98}.

\begin{figure} 
\begin{center}

\includegraphics[scale=0.7]{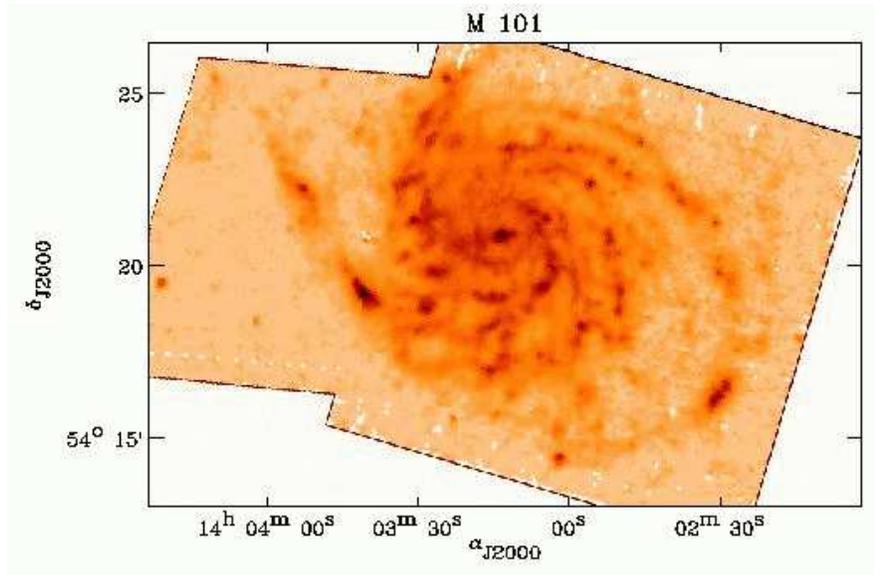}
\end{center}

\caption[]{The 6.75\,\mic\ map of M\,101, from
\inlinecite{rouatl}. The different features of the MIR morphology,
central region, \hii\ regions, spiral arms, are clearly seen. Diffuse
interarm emission is also present but it becomes rapidly too faint to
be detected.}

\label{fig:f2}
\end{figure}

A better grasp of the MIR morphology of spiral galaxies can be
obtained from studies that completely mapped their targets. Such
studies were predominantly made in the 6.75 and
15\,\mic\ broad band filters of ISOCAM (occasionally the ISOCAM 12\,\mic\
filter
was used). These filters were chosen to trace different components of the MIR
emission in galaxies. The 6.7\,\mic\ filter includes the most prominent
spectral features emitted by Polycyclic Aromatic Hydrocarbons (PAHs) powered
by UV photons, as revealed by spectroscopic studies of galactic sources. The
15\,\mic\ filter was originally chosen to trace stochastic
emission from very small grains, though it subsequently turned out that PAH
emission from the 12.7\,\mic\ feature and an underlying PAH continuum can also
contribute and even dominate the signal seen towards the spiral
arms. Morphological studies made using the above mentioned filters were
presented by \inlinecite{mal96} for
NGC\,6946, by
\inlinecite{smi98} for NGC\,7331, by \inlinecite{sau96} for M\,51 as
well as by \inlinecite{ben02a}, \inlinecite{dal00} and
\inlinecite{rouatl} for larger samples of spiral galaxies.
These papers show that, in the MIR, spiral galaxies present a
morphology that is quite similar to that observed at other
optical/NIR wavelengths. The spiral arms are very prominent, with
the giant \hii\
region complexes showing up as emission enhancements along the arms. A
diffuse underlying MIR emitting disk is often detected in the interarm
regions of the inner disk.
Finally, a central region of varying importance is observed. A clear
example of these morphological features is given in
Fig.~\ref{fig:f2} where the 6.75\,\mic\ map of M\,101 from \inlinecite{rouatl}
is presented.  Although the central region of spiral galaxies is
generally dominated optically by the bulge of old stars, one should
not jump to the conclusion that the central MIR component is dominated
by the Rayleigh-Jeans emission of cold photospheres or circumstellar
envelopes. The extent of the central MIR source is generally
different from that of the stellar bulge and furthermore
\inlinecite{roubar} showed that its IR colours are generally not
those expected from stars (i.e. the 15/6.75\,\mic\ flux ratio is
generally larger than unity whereas the opposite applies in
early-type galaxies where the stars provide most of the
6.75\,\mic\  emission, see
\opencite{ath02} or \opencite{xil04}, and
Fig.~\ref{fig:f9}). In fact, most of the emission from the
central regions of spiral galaxies can be attributed to dust emission
  powered by enhanced star
formation \cite{roubar}.

A further detailed study of the morphological features in the MIR is the
multiwavelength study of M\,83 by \inlinecite{vog04} who present the
azimuthally averaged radial profiles of a number of emission components of the
galaxy. These profiles show that the MIR emission of M\,83 can be
decomposed into a central component, associated with the central
starburst, the prominent arm structure and a diffuse exponentially
fading disk emission. The arm/interarm
contrast was found to be larger in the MIR than it is in the optical, although
it was found to be smaller than what is observed in \hal. This was interpreted
as being again due to the important, if not dominant role of star-formation
in powering the MIR emission, through the generation of UV
photons. The fact that
particles responsible for the MIR emission can be excited by a broad
range of photon energies (see e.g. \opencite{uch00} and
\opencite{li02}) explains why the influence of star-forming regions is
seen on broader scales than the actual extent of \hii\ regions (see
also Sect.~\ref{sssec:nirmirsed}).

An obvious question arising from these investigations pertains to the
colour of the various MIR emission components
(central regions, spiral arms, interarm regions), which could
potentially be used to reveal the heating sources of MIR
emission. Searches made in this direction came up with mixed results. For
instance
\inlinecite{sau96} reported a systematic variation of the 15/6.75\,\mic\ flux
ratio along some cuts through the arms of M\,51, while
\inlinecite{hel96} showed that, on average, the 15/6.75\,\mic\ flux ratio is
the same for the arm and interarm regions of NGC\,6946, and
\inlinecite{smi98} reported no systematic spectral variations in the
disk of NGC\,7331. The only systematic colour variations are observed
in the central regions of spiral galaxies, where the 15/6.75\,\mic\ flux
ratio can reach values observed in starburst galaxies
\cite{roubar}.
Another study of the physical interpretation of
the MIR morphology of spiral galaxies was presented
by \inlinecite{dal99}. This shows that although star-formation
activity is the primary driver for surface brightness variations
inside galaxies, it is only for regions of intense activity that a
colour variation occurs. Column density also appears as a secondary
parameter to explain the difference in disk surface brightness between
galaxies.

This relatively fast exploration of the IR morphology of spiral
galaxies thus reveals how the same processes that affect their visible
shapes, i.e. intensity and distribution of star-forming regions, play
an important part in determining their IR aspect. It already
suggests that the MIR and FIR part of the spectrum behave differently,
in the sense that spectral modifications are more apparent in the FIR
than in the MIR. This is not unexpected given the different
thermodynamical states of the grains that produce these emissions,
thermal equilibrium for most of the FIR, and stochastic heating for
most of the MIR.

\subsubsection{The extent of spiral disks in the FIR}
\label{sssec:firextent}

 Further information about the true distribution of dust in spiral
disks is provided by FIR observations of galaxies more distant
than the highly resolved local galaxies discussed in
Sect.~\ref{sssec:firmorph}, but still close enough to resolve the
diffuse disk at the longest FIR wavelengths accessible to ISO. In
a study of eight spiral galaxies mapped by ISOPHOT at
200\,${\mu}$m, Alton et al. (1998; see also Davies et al. 1999 for
NGC~6946) showed that the observed scalelength of FIR emission at
200\,${\mu}$m is greater than that found by IRAS at 60 and
100\,${\mu}$m.  Thus, the scalelength of the FIR emission
increases with increasing FIR wavelength. This result was
reinforced using LWS (\opencite{pec96}; \opencite{cgr03})
measurements of the dust continuum by Trewhella et al. (2000), and
can also be inferred from Fig.~2 of Hippelein et al. (2003) for
M~33. We note here that this implies that the bulk of the
200\,${\mu}$m emission arises from grains heated by a radially
decreasing radiation field, as would be expected for grains in the
diffuse disk. If most of the 200\,${\mu}$m emission had arisen
from localised sources associated with the parent molecular clouds
within the spiral arms, there should be no FIR colour gradient in
the galaxy, since the SEDs of the localised sources should not
depend strongly on position.

The second result to come out of the studies by Alton et al. (1998)
and Davies et al. (1999) is that the observed scalelength at
200\,${\mu}$m is comparable to or exceeds the scalelength of the
optical emission (see also Tuffs et al. 1996). As noted by Alton et al., this
result implies that
the {\it intrinsic} scalelength of the dust in galaxies is greater
than that of the stars. This is because the apparent scalelength of
stars should increase with increasing disk opacity (since the inner
disk is expected to be more opaque than the outer disk) whereas the
apparent scalelength of the dust emission will be less than the
intrinsic scalelength (due to the decrease in grain temperature with
increasing galactocentric radius).
The extraction of the precise relation between the
intrinsic scalelengths of stars and dust requires a self-consistent calculation
of the transfer of radiation through the disk (see Sect.~\ref{ssec:quant}).
The reason for the difference between the intrinsic scalelength of stars and
dust in galaxies is not self-evident, since
it is the stars themselves which are thought to be the sources of
interstellar grains (produced either in the winds of evolved
intermediate mass stars or perhaps in supernovae). One might speculate
either that there is a mechanism to transport grains from the inner
disk to the outer disk, or that the typical lifetimes of grains
against destruction by shocks is longer in the outer disk than it is
in the inner disk.

\begin{figure}[htb] 
\includegraphics[scale=0.7]{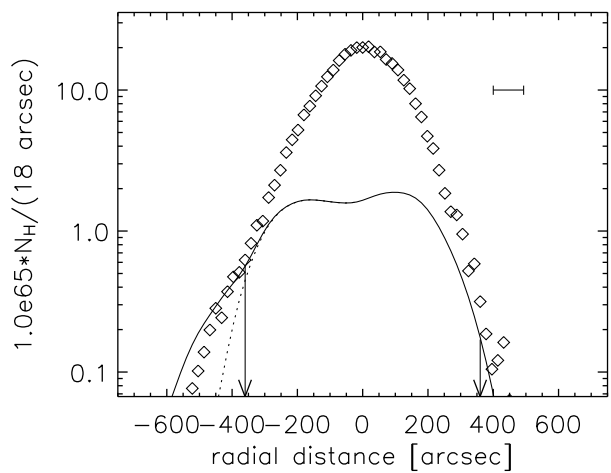}

  \caption[]{The radial profiles of \hi\ emission (from Swaters et al. 1997)
 convolved with the
  ISOPHOT PSF (solid line) and of $200\,{\mu}$m FIR emission (symbols)
  of NGC~891 (Popescu \& Tuffs 2003). Note that the extent and asymmetry of the
  200\,\mic\ emission follow that of the HI emission.
  The profiles are sampled at
  intervals of
  18$^{\prime\prime}$. The negative radii correspond to the southern side of
  the
  galaxy and the galaxy was scanned at 60 degrees with respect to the major
  axis. The units of the FIR profile are W/Hz/pixel, multiplied by a
  factor of $2\times 10^{-22}$ and the error bars are smaller than the
  symbols. The horizontal bar delineates the FWHM of the ISOPHOT
  PSF of 93$^{\prime\prime}$.
  The vertical arrows indicate the maximum extent of the optically
  emitting disk. The dotted line represents a modified \hi\ profile
  obtained in the southern side from the original one by cutting
  off its emission at the edge of the optical disk and by convolving
  it with the ISOPHOT PSF.}

\label{fig:f3}
\end{figure}

While Alton et al. and Davies et al. showed that the scalelength of
the 200\,${\mu}$m emission was comparable to or slightly larger than
that of the optical emission, these studies did not actually detect
grain emission beyond the edge of the optical disk. Since spiral
galaxies in the local universe are commonly observed to be embedded in
extended disks of neutral hydrogen - the so called ``extended \hi\
disks'', it is a natural question to ask whether these gaseous disks
contain grains.
This question was answered in the affirmative by
Popescu \& Tuffs (2003), through dedicated deep FIR maps of a large
field encompassing the entire \hi\ disk of the edge-on spiral galaxy
NGC~891, made using ISOPHOT at 170 and 200\,${\mu}$m (see Fig.~\ref{fig:f3}).

The large amounts of grains found in the extended \hi\ disk
(gas-to-dust ratio of $\sim 1\%$) clearly shows that this gaseous disk
is not primordial, left over from the epoch of galaxy formation. It
was suggested that the detected grains could have either been
transported from the optical disk (via the halo, using mechanisms such
as those proposed by Ferrara (1991), Davies et al. (1998), Popescu et
al. (2000a) or through the action of macro turbulence) or that they could
have been produced outside the galaxy (for example transferred in
interactions with other galaxies). It is interesting to note that,
although the dust emission is seen towards the \hi\ component, the
grains may not actually be embedded in the neutral ISM. Instead, this
dust could trace an ``unseen'' molecular component, as proposed by
Pfenniger \& Combes (1994), Pfenniger, Combes \& Martinet (1994),
Gerhard \& Silk (1996), and \inlinecite{val99b}.  This cold molecular gas
component has been invoked as a dark matter component to explain the
flat rotation curves of spiral galaxies. Its presence might also
reconcile the apparent discrepancy between the very low metallicities
measured in \hii\ regions in the outer disk (Ferguson, Gallagher \&
Wyse 1998) and the high ratio of dust to gas (on the assumption that all
gas is in form of HI) found by Popescu \&
Tuffs (2003) in the extended \hi\ disk of NGC~891.

\subsubsection{The extent of spiral disks in the MIR}
\label{sssec:mirextent}

As alluded to in the discussion of the MIR aspect of M\,31 above, the
MIR disks of spiral galaxies are generally quite short. At
first glance they can even appear truncated since in the outer region
the emission is generally dominated by the \hii\ regions of the arms,
superimposed on a very diffuse disk emission. This appearance is
however mostly an ``optical'' illusion. The radial profiles of the
exponentially declining MIR disk
emission of M\,83 \inlinecite{vog04} show a scalelength that
is clearly shorter than that seen in any optical bands, but with no
truncation. The strong decrease
of star-formation activity in the outer regions of
galaxies induces a rapid disappearance of the MIR emission, although,
as is definitely revealed by long-wavelength observations, dust is
still present at large galactocentric distances.

A more systematic study of the extent of MIR spiral disks was
presented in \inlinecite{rouatl}. These authors have defined a MIR
disk diameter, D$_{\rm MIR}$, in a similar way as the definition used
in the optical, by measuring the diameter of the 5\,${\mu}$Jy/arcsec$^{-2}$
isophote at 6.75\,\mic\ (this is roughly the detection limit of this
band). This definition allows to compare optical and MIR diameters on
a broad range of galaxies. It shows that the IR diameter (as defined
  above) is
systematically smaller than the optical one (the largest value of
D$_{\rm MIR}$/D$_{\rm opt}$ observed in this sample of about 70
galaxies is slightly less than 1, while the smallest value observed is
$\sim$0.3). The variation of D$_{\rm MIR}$/D$_{\rm opt}$ can be
related to both the morphological type of the galaxies or their \hi\
deficiency in the sense that earlier type or more severely \hi\
deficient galaxies have a smaller D$_{\rm MIR}$/D$_{\rm
opt}$. \inlinecite{rouatl} indicate that the \hi\ deficiency is likely
the dominant parameter given that \hi-deficient galaxies are
often classified as early-type galaxies, due to the absence of
star-forming regions at large galactocentric distances. The MIR
diameter of a galaxy is thus essentially determined by the extent of
the star-forming activity through its disk, demonstrating the crucial
role of star-formation in powering the MIR emission.

The fact that the MIR disks appear so much smaller than the FIR disks
immediately prompts the following question: does the dust responsible
for the MIR emission disappear at large galactocentric distances, or
phrased differently, should we observe a MIR counterpart to the FIR
extended distributions?
As mentioned before, ISOCAM was
probably less sensitive than ISOPHOT to faint diffuse emission, which would
bias the ISOCAM maps toward small-scale bright structures such as the
\hii\ regions in the disks. This would explain why MIR emission is not detected
far out in the disk, but still leaves open the question of why the
scalelength of the MIR emission is so much shorter than that of the FIR
emission.

\subsubsection{Comparison with morphologies at other wavelengths}
\label{sssec:compwave}

\begin{figure}[htb] 
\includegraphics[height=.3\textheight]{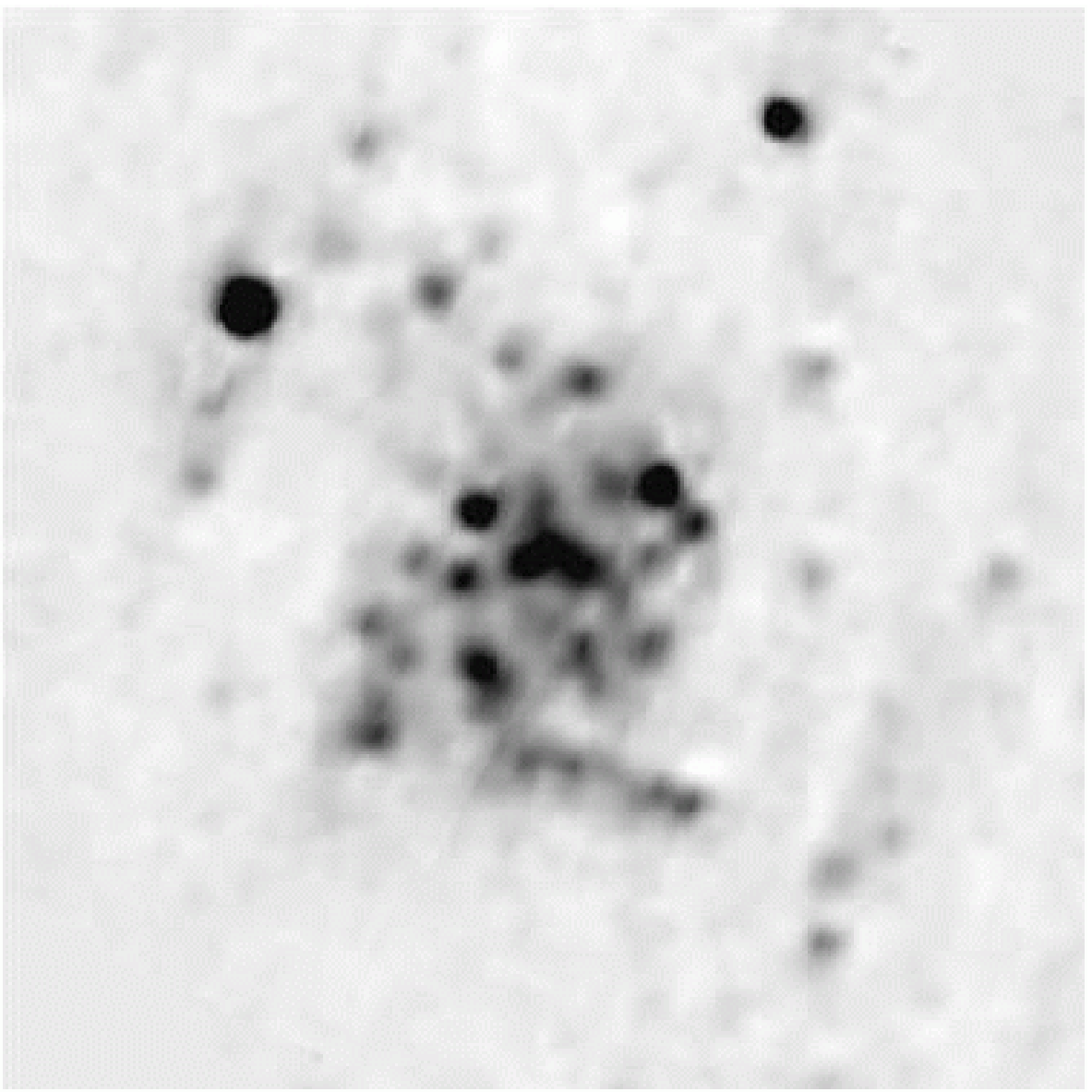}
\includegraphics[height=.3\textheight]{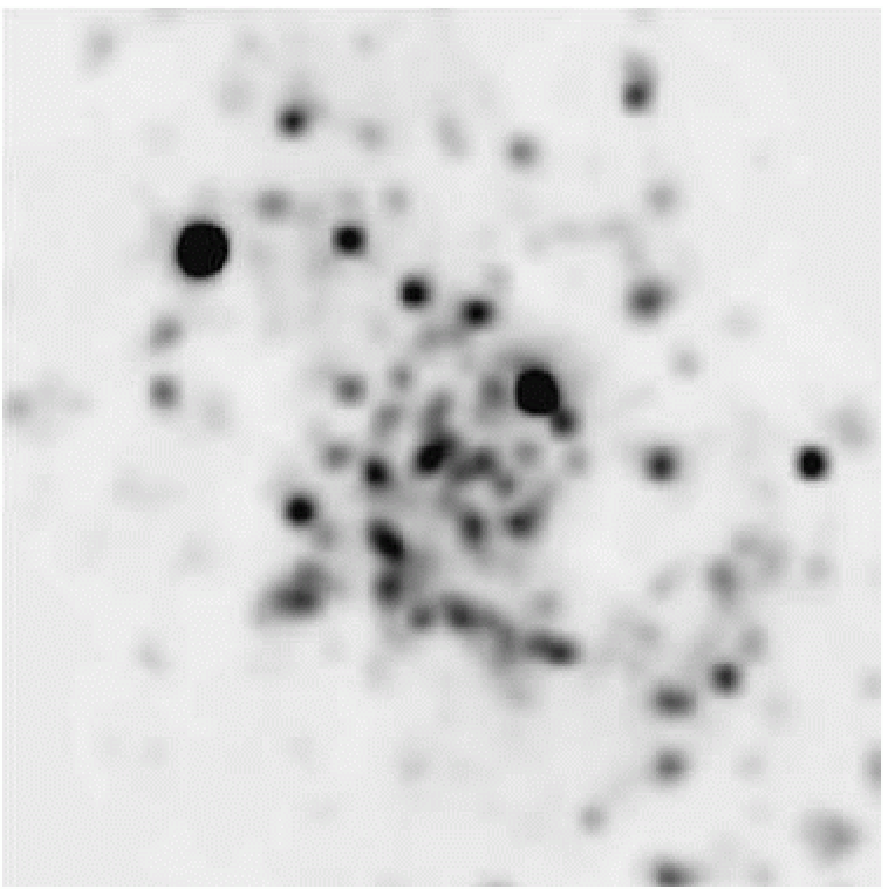}
\caption[]{Left: Distribution of the localised warm dust component at
  60\,${\mu}$m, $F^{\rm l}_{60}$, in M~33
  (Hippelein et al 2003). This is the scaled difference map
  $2(F_{60}-0.165\times F_{160}$), with the factor 0.165 given by the average
  flux density ratio $F_{60}/F_{170}$ in the interarm regions. Right:
  $H{\alpha}$ map of M~33 convolved to a resolution of 60$^{\prime\prime}$.}

  \label{fig:f4}
\end{figure}

\vspace{0.5cm}

\noindent
{\it Comparison with the optical}

\noindent
Although one immediately associates IR light with dust, the fact
that ISO could reach relatively short IR wavelengths prompts the
question of the contribution of stars to the shortest ISO bands. And
indeed, when studied quantitatively through the use of light
concentration indices,  as done for instance by \inlinecite{bos03b}, the
morphology of galaxies at MIR and NIR wavelengths show
surprising similarities. This indicates at least a similar spatial
distribution of dust and stars, which is not too surprising given the
origin of the MIR emission (see later). However, determining the actual
contribution of stellar photospheres and circumstellar shells to the
ISO fluxes is harder, as it requires first the construction of a pure
stellar image and second the correct extrapolation in wavelength
of this image to the
ISO wavelengths, taking into account the possible stellar population
gradients in the galaxy. This type of work led \inlinecite{bos03a} to
conclude that a large fraction of the 6.75\,\mic\ flux in galaxies
could be due to stars, ranging from 80\,\% in Sa to 20\,\% in Sc
galaxies. This conclusion is however not supported by the analysis of
\inlinecite{lu03} who showed that already at 6\,\mic, the mean flux
of their galaxies is one order of magnitude larger than what is
observed in galaxies devoid of interstellar medium (see
Fig.~\ref{fig:f9}). Therefore it appears unlikely that stars make
a large contribution to the IR emission of spiral galaxies
longwards of 6\,\mic, even though the spatial distribution of the emission
shares some morphological properties with that of the stars, a fact
which is due to their common link with the star-formation process.

\vspace{0.5cm}

\noindent
{\it Comparison with H${\alpha}$}

\begin{figure}[htb] 
\includegraphics[scale=0.38]{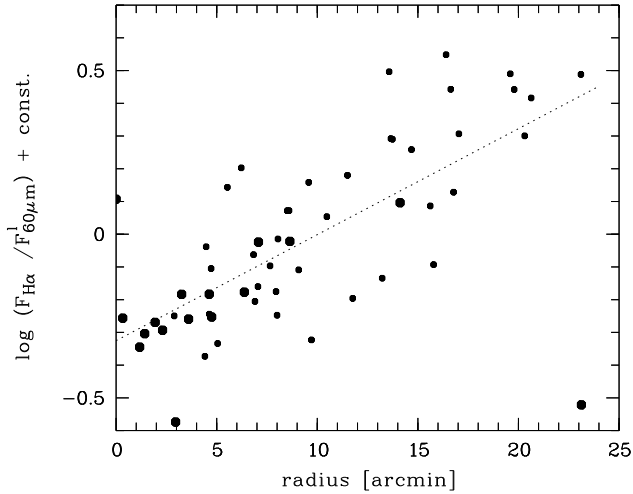}

\caption[]{The ratio of $F_{H\alpha}$ to the localised warm dust
component at 60\,${\mu}$m, $F^{\rm l}_{60}$, for the star-forming
regions in M~33 (Hippelein et al. 2003) versus distance from the
galaxy centre. Symbol sizes indicate the brightness of the sources.}

\label{fig:f5}
\end{figure}

\noindent
For the highly resolved galaxy M~33, Hippelein et al. (2003) showed
that there is a strong resemblance between the morphology of the
localised warm dust component at 60\,${\mu}$m ($F^{\rm l}_{60}$; obtained at
each direction by subtracting the 170\,\mic\ map scaled by the ratio of the
60/170\,\mic\ brightness in the interarm regions) and
the morphology of the H$\alpha$ emission (see Fig.~\ref{fig:f4}),
indicating that the 60\,${\mu}$m localised emission traces the
star-formation complexes.  The $F_{H\alpha}/F^{\rm l}_{60}$ ratio (see
Fig.~\ref{fig:f5}) for the star-formation complexes shows a clear
systematic increase with increasing radial distance from the centre
(allowing for the [NII] line contribution decreasing with distance,
the slope would be even steeper).  Very probably this is due to the presence
of a larger scale gradient of opacity affecting the recombination line
fluxes.

Similarly, in the MIR a number of authors have noted a
striking resemblance between MIR and \hal\ maps of galaxies, which
should not come as a surprise given the numerous links between MIR
emission and star-formation mentioned above. This is both the case for
the well studied galaxies M\,51 \cite{sau96}, NGC\,7331 \cite{smi98},
M~83 \cite{vog04} and NGC~6946 \cite{wal02} and for
larger samples of more distant objects (see for instance
\opencite{con01} or \opencite{dom03}). However, few authors have
studied the
relation between the two emissions {\em within}
galaxies. \inlinecite{sau96} presented a comparison of the MIR/\hal\
ratio for a number of \hii\ region complexes in M\,51. This reveals
that extinction affecting the measure of \hal\ is responsible for most
of the variation observed in the MIR/\hal\ ratio.

One should note that the apparently strong spatial correlation between
the MIR and \hal\ maps is slightly misleading, in the sense that the
limited spatial resolution tends to make the maps more similar than
they really are. Studies of resolved \hii\ regions indeed show that
the MIR emission tends to avoid the peaks of \hal\ emission and rather
delineates the edges of molecular clouds exposed to the radiation
produced by newly-formed stars (see e.g. \opencite{con98} or
\opencite{con00} and Fig.~\ref{fig:f8}).

The complexity of the local relation between the MIR emission and
\hal\ is revealed by the studies of \inlinecite{vog04} and
\inlinecite{wal02}. Using maps at the same resolution, and only taking
independent points, these authors showed that the local MIR-\hal\
correlation is quite poor and highly non-linear. As we will see, this
is in contrast to the global MIR-\hal\ correlation, which is linear
and tighter. A plausible explanation of this, first introduced in the context
of the FIR emission by Popescu et al. (2002; see also Pierini et al. 2003b),
would be the existence of a diffuse component of the MIR emission from the
underlying disk, powered by the star-formation activity, but
on larger scales than that used by \inlinecite{wal02} or
\inlinecite{vog04}.

\vspace{0.5cm}
\noindent
{\it Comparison with UV}

\begin{figure}[htb] 
\includegraphics[scale=0.66]{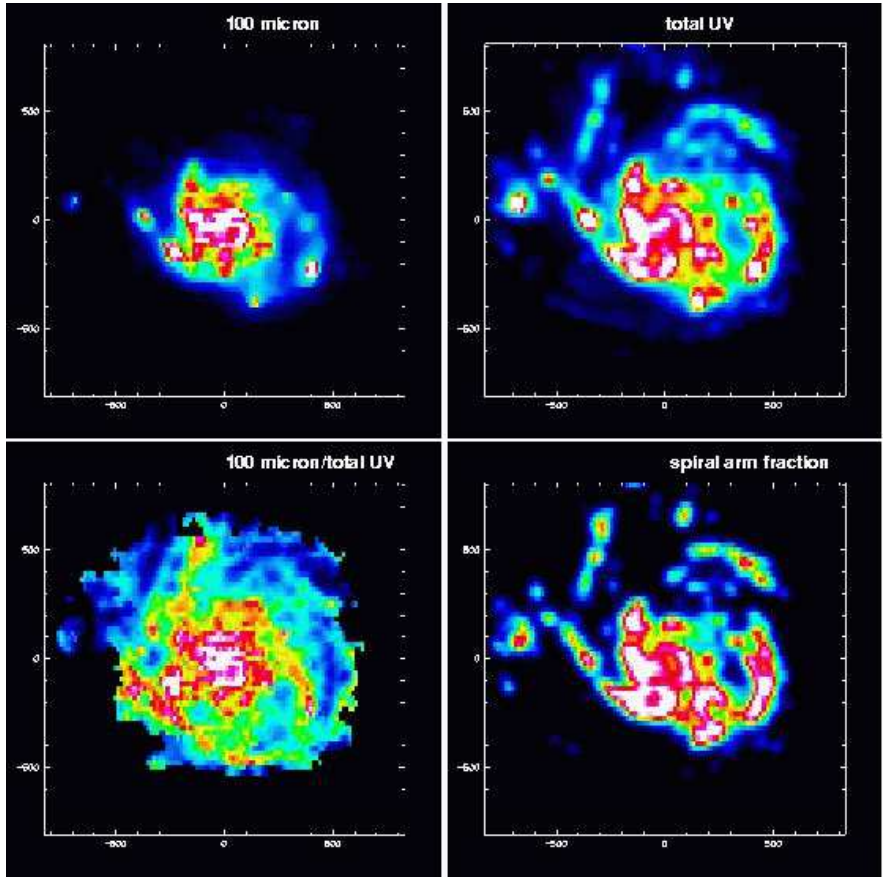}

\caption{FIR-UV comparison for M~101 (Popescu et al. 2005). Top left
panel: filter-integrated 100\,${\mu}$m ISOPHOT image.  Top right
panel: ``total UV'' image converted to the orientation, resolution and
sampling of the 100\,${\mu}$m ISOPHOT image. Bottom left panel: the
ratio image of the filter-integrated 100\,${\mu}$m ISOPHOT image
divided by the corresponding ``total UV'' image. Bottom right panel:
the image of the ``spiral arm fraction'' at the orientation,
resolution and sampling of the 100\,${\mu}$m ISOPHOT image.  All
panels depict a field of 27.7$^{\prime}\times 27.1^{\prime}$ centered
at $\alpha^{2000}=14^{\rm h}03^{\rm m}13.11^{\rm s}$;
$\delta^{2000}=54^{\circ}21^{\prime}06.6^{\prime\prime}$. The pixel
size is $15.33^{\prime\prime}\times 23.00^{\prime\prime}$.}

\label{fig:f6}
\end{figure}

A fundamental property of spiral galaxies is the fraction of light
from young stars which is re-radiated by dust. This property can be
investigated as a function of position in the galaxy by a direct
comparison of ISOPHOT maps at 60, 100 and 170\,${\mu}$m with UV maps
obtained with GALEX (Galaxy Evolution Explorer; Martin et al. 2005) in
its near-UV (NUV; 2310\,\AA) and far-UV (FUV; 1530\,\AA) bands. Such a
comparison was performed for M~101 by Popescu et al. (2005).  The top
panels in Fig.~\ref{fig:f6} display the 100\,${\mu}$m ISOPHOT image
(left) together with the corresponding ``total UV'' (integrated from
1412 to 2718\,\AA) image (right).
Comparison between the ratio image
(100\,${\mu}$m/UV) (bottom left panel) and an image of the ``spiral
arm fraction'' (the fraction of the UV emission from the spiral arm within an
ISOPHOT beam; bottom right panel) shows that the high values of the
100\,${\mu}$m/UV ratio trace the interarm regions. In other words the
``spiral features'' in the ratio image are in reality regions of
diffuse emission which are interspaced with the real spiral features,
as seen in the ``spiral arm fraction'' image.

\begin{figure}[htb] 
\includegraphics[scale=0.6]{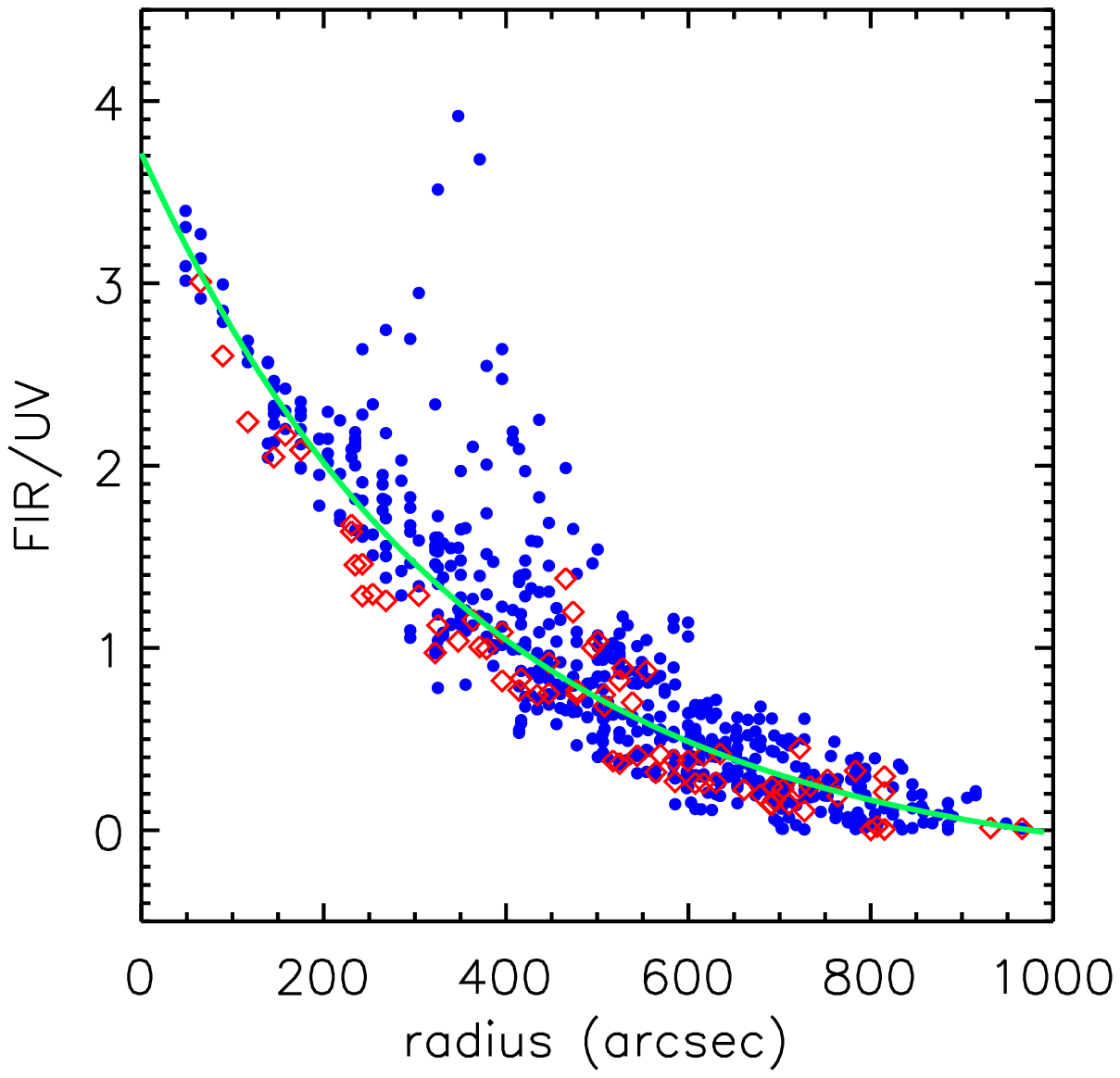}

\caption{The pixel values of the FIR/UV ratio map of M~101 (Popescu et
al. 2005) at the resolution of the 170\,${\mu}$m image versus angular
radius. The blue dots are for lines of sight towards interarm regions
and the red diamonds towards the spiral arm regions. The green solid
line is an offset exponential fit to the data.}

\label{fig:f7}
\end{figure}

The trend for the FIR/UV ratio to be higher in the
diffuse interarm regions than in the spiral-arms is seen in
Fig.~\ref{fig:f7} from the segregation of the blue dots and red
diamonds at a given radius. This apparently surprising result was
explained in terms of the escape probability of UV photons from spiral
arms and their subsequent scattering in the interarm regions, and in
terms of the larger relative contribution of optical photons to the
heating of the dust in the interarm regions. The combined effect of
the optical heating and the scattering of the UV emission means that
the FIR/UV ratio will not be a good indicator of extinction in the
interarm region.

Despite these local variations, the main result of Popescu et
al. (2005) is the discovery of a tight dependence of the FIR/UV ratio
on radius, with values monotonically decreasing from $\sim 4$ in the
nuclear region to nearly zero towards the edge of the optical disk
(see Fig.~\ref{fig:f7}).  This was interpreted in terms of the
presence of a large-scale distribution of diffuse dust having a
face-on optical depth which decreases with radius and which dominates
over the more localised variations in opacity between the arm and
interarm regions.

A comparison of the UV and MIR morphologies, analogous to the UV-FIR
  comparison for M~101, has been made
  for  the south-west ring of M\,31 by
\inlinecite{pag99}. Here the 20$''$ resolution 200\,nm map from the
FOCA~1000 balloon experiment is compared to the $\sim$ 6$''$
resolution 6.75\,\mic\ ISOCAM map (Fig.~\ref{fig:f8}). This study,
and especially its Fig.~8, shows that although the overall morphology
of the two maps appears similar, following the ring-like structure
observed in that region of the galaxy, the UV and MIR emission
actually complement each other: UV-emitting regions fill the holes in
the distribution of the 6.75\,\mic\ emission. In fact, the same study
shows a much tighter spatial correlation of the MIR emission with the
gaseous components (\hi\ or CO) than with the UV emission. Again, this
is interpreted as showing that even if most of the energy
that ultimately powers the MIR emission is generated in the
giant \hii\ region complexes that are seen in \hal\ or UV light, most
of the MIR emission comes from the surfaces of the molecular clouds
that {\em surround} these complexes.

\begin{figure} 
\begin{center}
\includegraphics[width=\textwidth]{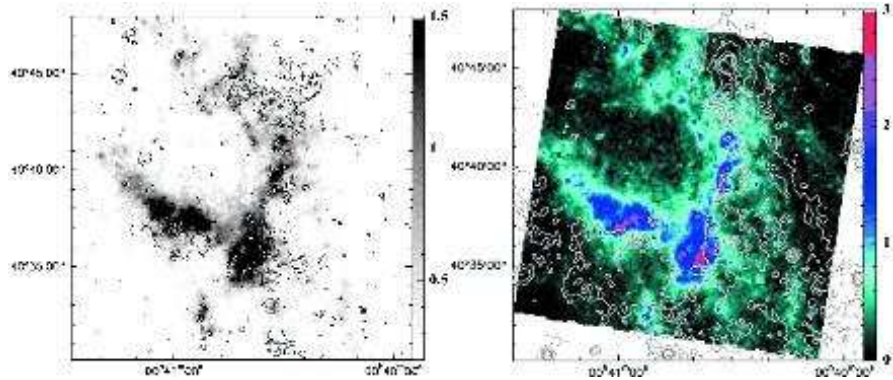}
\end{center}
\caption[]{The SW ring of M\,31 observed by \inlinecite{pag99}. On the
left panel, \hal\ contours in black are superimposed on a grey scale
map at 6.75\,\mic. On the right panel, UV contours are superposed on
the same map, in colour this time. In both maps, though the overall
ring-like structure is preserved at all wavelengths, it is clear that
the UV and \hal\ emissions tend to be located in regions of minimal
6.75\,\mic\ emission, and that reversely, the MIR emission is located
on the edges of the \hal\ and UV regions.}

\label{fig:f8}
\end{figure}

These differences in the UV and MIR spatial distribution are confirmed
by the recent observations of the Spitzer satellite,
which also reveal that at 24\,\mic\
the association between IR
and UV or \hal\ emission becomes tighter, even at high spatial
resolution \cite{hel04}. The reason for this will be elucidated in
section~\ref{sssec:nirmirsed}.

\vspace{0.5cm}
\noindent

\subsection{Integrated properties}
\label{ssec:int}

The overriding result of all ISOPHOT studies of the integrated properties
of normal galaxies in the FIR is that their SEDs in the
40-200\,${\mu}$m spectral range require both warm and cold dust
emission components to be fitted. Although the concept of warm and cold
emission components is as old as IRAS (de Jong et al. 1984), it only became
possible to directly measure and
spectrally separate these components using ISOPHOT's multi-filter coverage
of the FIR regime out to 200\,${\mu}$m.

In the MIR the two most stunning results are the
high similarity of the disk's SED in spiral galaxies and the almost complete
invariance of the global MIR colours to the actual star-formation
activity of the galaxy as a whole. The former result shows that the MIR
emission comes predominantly from PDRs, whereas the latter result demonstrates
the importance of the emission from the central region to the
MIR, a characteristic which could only be uncovered due to the high spatial
resolution offered by ISOCAM.

In order to investigate the integrated properties of local universe gas-rich
galaxies, a number of statistical samples were constructed, which were
primarily observed by ISOPHOT and ISOCAM.
All these projects were complementary in terms of selection and observational
goals. In descending order of depth (measured in terms of a typical
bolometric luminosity of the detected objects), the published surveys are:

{\bf The ISO Virgo Cluster Survey} is a combination of the
ISOPHOT Virgo Cluster Deep Survey (IVCDS; Tuffs et al. 2002a,b;
Popescu et al. 2002) and ISOCAM measurements (\opencite{bos03b}, see also
\opencite{bos97}) of the same sample supplemented by additional targets on
the eastern side of
the cluster. The IVCDS represents the {\it deepest survey} (both in
luminosity and surface brightness terms) of normal galaxies
measured in the FIR with ISO. A complete volume- and luminosity-limited
sample of 63
gas-rich Virgo Cluster galaxies selected from the Virgo Cluster
Catalogue (Binggeli et al. 1985; see also Binggeli et al. 1993) with
Hubble types later than S0 and brighter than $B_{\rm T} \le 16.8$ were
mapped with ISOPHOT at 60, 100 and 170\,${\mu}$m.
A total of 99 galaxies were mapped with ISOCAM in the 6.7 and 15\,\mic\
filters.
The IVCDS sample was (in part) also observed with the LWS (Leech et
al. 1999).

The ISO Virgo Cluster Survey provides a database for statistical
investigations of the FIR and MIR SEDs of gas-rich
galaxies in the local universe spanning a broad range in star-formation
activity and morphological types, including dwarf systems and galaxies with
rather quiescent star-formation activity.

{\bf The Coma/A1367 Survey} \cite{con01} consists of 6 spiral and 12
irregular galaxies having IRAS detections at 60$\,{\mu}$m. The
galaxies were selected to be located within 2 or 1 degrees of the
X-ray centres of Coma and A1367 clusters, respectively, with emphasis
on peculiar optical morphologies. Each galaxy was observed in a single
pointing with ISOPHOT,
at 120, 170 and 200$\,{\mu}$m, as well
as mapped with ISOCAM in the 6.75 and 15\,\mic\ broadband filters. The sample
provides a database of integrated flux densities for a
pure cluster sample of high luminosity spiral and irregular galaxies.

{\bf The ISO Bright Spiral Galaxies Survey} (Bendo et al. 2002a,b;
2003) consists of 77 spiral and S0
galaxies chosen from the Revised Shapley-Ames Catalog (RSA), with
$B_{\rm T}\,\le\,12.0$. Almost all are IRAS sources. Mainly an ISOCAM
mapping survey with the 12\,\mic\ filter, the
project also used ISOPHOT to take 60, 100 and 170$\,{\mu}$m short
stares towards the nucleus of the galaxies and towards background
fields. The sample provides a database of MIR morphologies and FIR surface
brightnesses of the central regions of bright spiral galaxies, including S0s.

The ISO Bright Spiral Galaxies Survey and the ISO Virgo Cluster Survey
represent
the principle investigations of optically selected samples of normal
galaxies. It should be emphasised that the main difference between
them is primarily one of shallow versus deep, rather than field versus
cluster, since by design the Virgo Sample predominantly consists of
infalling galaxies from the field, and no cluster specific effects
could be found (see also Contursi et al. for Coma/A1367 Sample).

{\bf The ISO Key Project on Normal Galaxies Sample} \cite{dal00}
consists of 69 galaxies selected to span the whole range of the
classical IRAS colour-colour diagram \cite{hel86}. Since IRAS detected a
vast number of galaxies in its four bands, the selection was also made
to span the Hubble sequence evenly and provide a broad range of
IR luminosities, dust temperatures (as determined by IRAS) and
star-formation activity. Galaxies in the sample were mapped
with ISOCAM, their main cooling lines in the FIR were measured with
ISOLWS \cite{mal01} and their 2.5-12\,\mic\ spectra were measured with
ISOPHOT-S \cite{lu03}.

{\bf The ISOCAM Parallel Mode Survey} \cite{ott03} is drawn from
observations made with ISOCAM, mostly around 6\,\mic, while another
ISO instrument was prime. It detected around 16000 distinct objects
down to 0.5\,mJy. Identifications of these objects by
cross-correlation with other catalogues is on-going and $\sim$25\% of
the objects are expected to be galaxies.

{\bf The ISOPHOT Serendipity Survey} (Stickel et al. 2000) has initially
catalogued 115 galaxies with $S_{\nu}\,\ge\,2$\,Jy at 170$\,{\mu}$m
and with morphological types predominantly S0/a\,-\,Scd. This sample
provides a database of integrated 170$\,{\mu}$m flux densities for
relatively high luminosity spiral galaxies, all detected by IRAS at 60
\& 100$\,{\mu}$m. Recently a catalogue of 1900 galaxies was released
(Stickel et al. 2004), of which a small fraction does not have IRAS
detections. Most of the 1900 galaxies are spirals. The
measured 170\,${\mu}$m flux densities range from just below 0.5\,Jy up
to $\sim 600$\,Jy.

Finally about 30 additional spiral galaxies were mapped
in various ISO guaranteed and open time programs (see
\opencite{rouatl}, for a compilation)

\subsubsection{The FIR spectral energy distribution: Dust temperatures,
masses and luminosities}
\label{sssect:firsed}

The presence of a cold
dust emission component peaking longwards of 120\,${\mu}$m was
inferred from studies of the integrated SEDs of individual
galaxies (see Sect.~\ref{sssec:firmorph}), from statistical studies of
small samples (Kr\"ugel et al. 1998; Siebenmorgen et
al. 1999) and was confirmed and generalised by studies of the larger
statistical samples mentioned above.
The latter studies also demonstrated the universality
of the cold dust component, showing it to be present within all types
of spirals (Tuffs \& Popescu 2003).
The cold emission component predominantly arises from dust
heated by the general diffuse interstellar medium and the warm component
from locally heated dust in HII regions, an interpretation consistent
with what has been seen in the ISOPHOT maps of nearby galaxies (see Sect.~\ref{sssec:firmorph})
and with self-consistent modelling of the UV-FIR SEDs (see Sect.~\ref{ssec:quant}).

The cold dust component is most prominent
in the most
``quiescent'' galaxies, like those contained in the IVCD sample, where
the cold dust temperatures were found to be broadly distributed, with
a median of 18\,K (Popescu et al. 2002), some $8-10$\,K lower than
would have been predicted by IRAS.
The corresponding dust masses were
correspondingly found to be increased by factors of typically $2-10$
(Stickel et al. 2000) for the Serendipity Sample and by factors $6-13$
(Popescu et al. 2002) for the IVCD sample, with respect to previous
IRAS determinations. As a consequence, the derived gas-to-dust ratios
are much closer to the canonical value of $\sim 160$ for the Milky Way
(Stickel et al. 2000, Contursi et al. 2001; see also Haas et al. 1998
for M~31), but with a broad distribution of values (Popescu et
al. 2002).

It was found that the cold dust component provides not only
the bulk of the dust masses, but even the bulk of the FIR luminosity,
in particular for the case of the most quiescent spirals, like those
in the IVCD sample. In contrast to the SEDs found by the other ISOPHOT
studies, which typically peaked at around 170\,${\mu}$m, Bendo et
al. (2003) derived spatially integrated SEDs typically
peaking at around 100\,${\mu}$m. The result of Bendo et al.  may
reflect the fact that these observations were single pointings, made
towards the nucleus of resolved galaxies
extending (in the main) beyond the field of view of ISOPHOT, and were therefore
biased towards nuclear emission, which is warmer than the extended
cold dust emission missed (or only partially covered) by these
measurements. Nevertheless, the measurements of Bendo et al. constitute a
useful probe of the FIR emission of the inner disks. This emission
(normalised to K band emission) was found to increase along the Hubble
sequence (Bendo et al. 2002b).

Since the FIR carries most of the dust luminosity, it is interesting to
  re-evaluate the question of the fraction of stellar photons converted via
  grains into IR photons, taking into account the comprehensive
  measurements of the cold dust emission component made available by
  ISOPHOT. This was done by
Popescu \& Tuffs (2002a), who showed that the mean percentage of stellar
light reradiated by dust is $\sim30\%$ for the Virgo Cluster
spirals contained in the IVCD sample. This study also included the dust
  emission radiated in the NIR-MIR range. The fact that the mean value of
$\sim30\%$ found for the Virgo Cluster spirals is the same as the canonical value
  obtained for the IRAS Bright Galaxy Sample (BGS; Soifer \& Neugebauer 1991)
  is at first sight strange, since IRAS was not sensitive to the cold dust
  component. However the BGS sample is an IR selected sample and
  biased towards galaxies with higher dust luminosities, while the Virgo
  sample is optically selected and contain a full representation of quiescent
  systems. So the deficit in FIR emission caused by sample selection criteria
  for the Virgo sample is compensated for by the inclusion of the cold dust
 component.
Popescu \& Tuffs (2002a) also found evidence for an
increase of the ratio of the dust emission to the total stellar
emitted output along the Hubble sequence, ranging from typical values
of $\sim 15\%$ for early spirals to up to $\sim 50\%$ for some late
spirals. This trend was confirmed by \inlinecite{bos03a} who further
utilised the new ISO data on dust emission to constrain the
corresponding absorption of starlight and thus improve extinction
corrections (using the technique pioneered by Xu \&
Buat 1995 for the IRAS data).

\subsubsection{The NIR to MIR spectral energy distribution}
\label{sssec:nirmirsed}

Using the large body of NIR-MIR data - both photometric and
spectroscopic - from the statistical studies
  mentioned previously, one can also determine the general properties
of the SED of galaxies from the NIR to the MIR.

\begin{figure} 
\includegraphics[width=\textwidth]{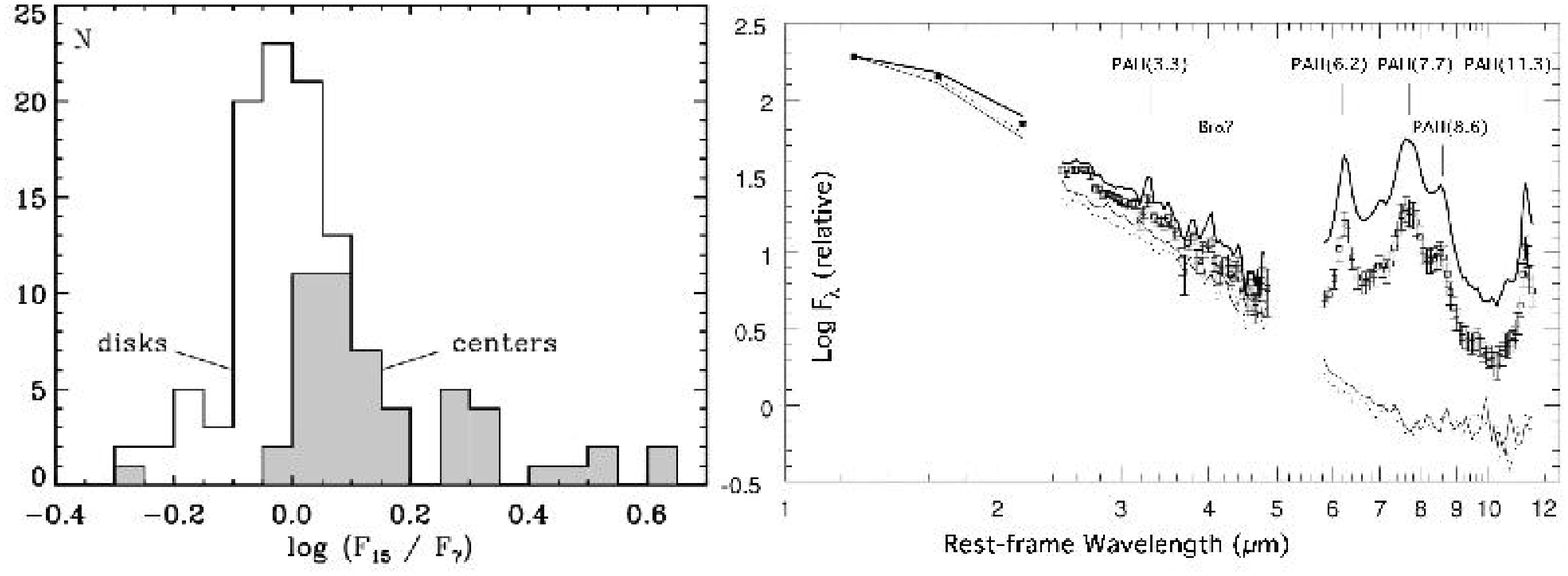}

\caption[]{On the left panel, we show the distribution of
15/6.75\,\mic\ flux ratio for the disks and central regions of spiral
galaxies (taken from \opencite{roubar}). It is clear that the disks
have a very uniform colour while the central regions explore a much
larger range. On the left side, the global NIR-MIR SED (from \opencite{lu03}) reveals the dominance of the PAH
features in the MIR, as well as the very small dispersion from one
object or the other. Both PAH dominated spectra represent the average
spectrum of the sample, with two different normalisation. The very
small dispersion bars reveal the homogeneity of the sample. The thin
solid and dotted lines are spectra from early-type galaxies, dominated
by stars.}

\label{fig:f9}
\end{figure}

Concerning
the colours of galaxies in the MIR, the main property
that is evidenced is a remarkable uniformity of the 6.75/15\,\mic\
flux ratio. In contrast with the well-known IRAS colour-colour diagram,
the composite ISO-IRAS colour diagram (showing the 6.75/15\,\mic\ flux
ratio versus the 60/100\,\mic\ flux ratio, see \opencite{dal00}) shows
a plateau of the 6.75/15\,\mic\ flux ratio for a large range of
60/100\,\mic\ flux ratios. A decrease of the ISO colour is only
observed for the largest values of the IRAS colour. A rapid
interpretation of that behaviour is that for most galaxies, the ISOCAM
filters collect emission from transiently heated PAHs
which shows no spectral dependence
on the heating intensity. It is only for the most actively
star-forming galaxies that a fraction of the 15\,\mic\ flux originates
in hot dust located in the \hii\ regions. This simple interpretation
has to be taken with caution since it assumes that the integrated
colour reflects a global property of the object (e.g. its star-forming
activity). Indeed, following up on the fact that in the MIR the
central regions of spiral galaxies appear to differ from the disk (see
Sect.~\ref{sssec:mirmorph}), \inlinecite{roubar} inspected the behaviour of
the 6.75/15\,\mic\ flux ratio separately for disks and central
regions. This revealed that the disk colours are extremely similar from
one galaxy to the other and that it is the fraction of the total flux
emitted from the central region combined with the colour of that region
that dominate the variation of the global 6.75/15\,\mic\ flux ratio,
rather than the global level of star-formation activity
(Fig.~\ref{fig:f9}). We remark that, in principle, the decrease of the
6.75/15\,\mic\ flux ratio in the
central regions could correspond to a drop of the 6.75\,\mic\ flux
rather than to an increase of the 15\,\mic\ flux. That this is not the
case is clearly demonstrated by the ISOCAM spectra of these central
regions \cite{roubar} that show the appearance of the expected
small grains continuum at the long wavelength edge of the bandpass.

This uniformity of the disk emission is confirmed by the spectroscopic
study of \inlinecite{lu03}. With the spectroscopic capabilities of
ISOPHOT-S, they showed that the 2.5-12\,\mic\ spectrum of spiral
galaxies is indeed extremely uniform. The NIR part is dominated by the
tail of photospheric emission, while the MIR part presents the
classical features of PAH emission, with extremely small spectral
variations of these features from one galaxy to the other (of the
order of 20\%, see Fig.~\ref{fig:f9}). Furthermore, these
variations are not related to the IRAS 60/100\,\mic\ colour or to the
FIR-to-blue luminosity ratio. Thus, as revealed by the broad-band
colours, it appears that the detailed SED of
spiral galaxies in the MIR show very little dependence on the actual
level of star-formation in the galaxy. Studies of the MIR SED of
various regions in our own galaxies have shown that a spectrum such as
that observed globally in spiral galaxies is prototypical of
photo-dissociation regions (PDRs), e.g. the outer layers of molecular clouds
exposed to moderately strong radiation fields (see
e.g. \opencite{van04}). Hence the spatial distribution of the MIR
emission revealed in Sect.~\ref{ssec:spatdist} is fully consistent
with the observed SED.

It should be noted, however, that the ISOPHOT-S spectral range is
shorter than the one explored by the 15\,\mic\ filter of ISOCAM, and
in fact covers a wavelength range that is not very sensitive
spectrally to activity variations, as demonstrated by the study of
starburst galaxies presented by \inlinecite{for03}. Therefore it is
from the combination of the nearly constant colours observed by
\inlinecite{roubar} and the constant spectrum observed by
\inlinecite{lu03} that one can conclude that the 5-18\,\mic\ spectrum
of galactic disks is dominated by the emission from PDRs, rather than from
the \hii\ regions themselves.

The systematic exploration of the NIR-MIR SED of spiral galaxies made
possible by ISO offered a number of other important discoveries. First
\inlinecite{lu03} evidenced a new component of dust emission
consisting of a ``hot'' NIR continuum emission. Its association with
the ISM was made clear by the existence of a correlation between that
NIR excess and the strength of the PAH features. However a physical
explanation of this continuum is still lacking. Second, the strength
of the PAH emission (which represents most, but not all of the MIR
flux when defined as the 5-20\,\mic\ luminosity for instance), was
found to be well correlated spatially with the 850\,\mic\ flux of bright
  regions inside
a number of galaxies \cite{haa02}.
This can be
interpreted as showing a close physical association between cold dust
clouds and the PAHs, but, given the limited spatial resolutions
involved, more likely indicates that the localised 850\,\mic\
emission comes essentially from the inside of the molecular clouds whose
surfaces produce the PAH features.
Finally \inlinecite{hel01} found a strong
correlation between the 5-10\,\mic\ luminosity and  the \cii\ line luminosities of
actively star-forming galaxies from the sample of Malhotra (2001). Since the
[CII] emission from this sample mainly originates in PDRs (see Sect.~2.4.1),
this correlation  implies that PAHs,
which dominate the 5-10\,\mic\ luminosities \cite{lu03}, are also
responsible for most of the gas heating in strongly star-forming
galaxies.

\subsubsection{Two outstanding questions of the IRAS era}
\label{sssect:quest}

\medskip
\noindent
{\it The radio-FIR correlation}

\begin{figure}[htb] 
\includegraphics[scale=0.40]{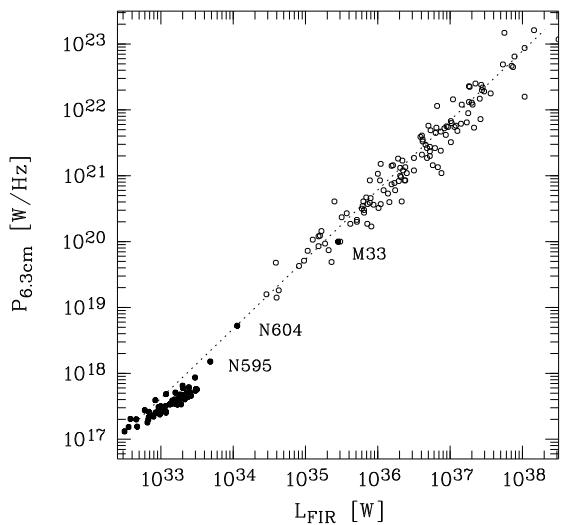}

\caption{Plot of the monochromatic radio luminosity versus the FIR
luminosities for M~33 (Hippelein et al. 2003) and its star-forming
regions (filled circles) together with the data for the Effelsberg
100-m galaxy sample (Wunderlich et al. 1987, open
circle). The dotted line has a slope of 1.10 (Wunderlich \& Klein
1988).}

\label{fig:f10}
\end{figure}

One of the most surprising discoveries of the IRAS all-sky survey was
the very tight and universal correlation between the spatially
integrated FIR and radio continuum emissions (de Jong et al. 1985; Helou et
al. 1985; Wunderlich et al. 1987; see V\"olk \& Xu 1994 for a review).
However all
the pre-ISO studies of the FIR/radio correlation were based on FIR
luminosities derived from the IRAS 60 and 100\,${\mu}$m flux
densities, and thus were missing the bulk of the cold dust
luminosity. The ISOPHOT measurements at 60, 100 and 170\,${\mu}$m were
used to redefined the FIR/radio correlation (Pierini et al. 2003b) for a
statistical sample of spiral galaxies. The inclusion of
the cold dust component was found to produce a tendency for the total
FIR/radio correlation to become more non-linear than inferred from the
IRAS
  60 and 100\,\mic\ observations. The use of the three FIR
wavelengths also meant that, for the first time, the correlation could
be directly derived for the warm and cold dust emission
components.
The cold FIR/radio correlation was found to be slightly
non-linear, whereas the warm FIR/radio correlation is linear. Because
the effect of disk opacity in galaxies would introduce a non-linearity in
the cold-FIR/radio correlation, in the opposite sense to that
observed, it was argued that both the radio and the FIR emissions are
likely to have a non-linear dependence on SFR. For the radio emission an
enhancement of the small free-free component with SFR can account for this
effect. For the cold FIR emission a detailed analysis of the dependence
of local absorption and opacity of the diffuse medium on SFR is required to
understand the non-linear trend of
the correlation (see Pierini et al. 2003b).

The improved angular resolution of ISO compared with IRAS also allowed a more
detailed examination of the local FIR-radio correlation on sub-kpc size
galactic substructures. Hippelein et al.
(2003) established the correlation for the
star-forming regions in M~33. This correlation is shown in
Fig~\ref{fig:f10}, overplotted on the correlation for integrated
emission from galaxies. It is apparent that the local correlation has
a shallower slope (of the order of 0.9) than for the global
correlation. It was argued that the local correlation is attributable to the
increase with SFR of dust absorption in increased dust densities, and to local
synchrotron emission from within supernova remnants, still confining their
accelerated electrons. Both emission components play only a minor role in the
well known global radio-FIR correlation, that depends on the dominant
large-scale absorption/re-emission properties of galaxies.

This local correlation has also been investigated in the MIR by
\inlinecite{vog04}. A good correlation was found between the two emissions
on scales of $\sim 500$\,pc, which is comparable to the scales of the
individual regions considered for the local FIR-radio correlation in
M~33.  The slope of the local MIR-radio correlation was comparable to
that of the local FIR-radio correlation. This is consistent with the MIR and
FIR emissions on scales of $\sim 500$\,pc being powered by the same UV
photons.

\medskip
\noindent
{\it Infrared emission as a star-formation tracer}

Because star-forming regions and \hii\ complexes shine
brightly in the IR, and because starburst galaxies output most of
their energy in the IR, this wavelength domain has always been
associated with star-formation. As we have seen earlier, this is
supported by a qualitative analysis of the IR emission of
galaxies, but a quantitative assessment of this association has proven
difficult. With the enhanced capacities of ISO, this question has been
addressed in much more detail, and this reveals that the different
regions of the IR SED have a different link with star-formation.

In the FIR, the correlation with the most widely-used star-formation
tracer, \hal, is extremely non-linear, which has long led authors to
suspect that the FIR emission is the result of more than one component
(see e.g. \opencite{lon87}). With ISOPHOT the correlation was separately established for the warm and cold dust emission components.
A good linear correlation was found between the
warm FIR luminosities (normalised to the K band luminosity) derived for the
IVCD sample and their H$\alpha$
EW (Popescu et al. 2002). This is in agreement with the assumption
that the warm dust component is mainly associated with dust locally
heated within star-formation complexes. The scatter in the correlation was
  attributed to a small component of warm emission from the diffuse
disk (produced either by transiently heated grains or by grains heated by
  old stellar population), as well as to the likely variation in HII region
dust temperatures within and between galaxies.
A good but non-linear
correlation was
found between the cold FIR luminosities of the
galaxies from the IVCD sample and their H$\alpha$ EW, in the sense that
FIR  increases more slowly than H$\alpha$. Since the bulk of the
cold FIR emission arises from the diffuse disk, the existence of this
correlation implies that the grains in the diffuse disk are mainly powered by
the UV photons (see also Sect.~\ref{ssec:quant}).
The non-linearity of the
correlation is consistent with there being a higher contribution from optical
photons in heating the grains in more quiescent galaxies.

For the late-type galaxies in the Coma and A1367 clusters, Contursi et
al. (2001) derived the relationships between the IR flux
densities at 200, 170, 120, 100 and 60\,\mic, normalised to the H band flux,
as a function of the H${\alpha}$ EW. It was found that the poorer
correlation is in the 200\,${\mu}$m band and that the values of the
fitted slopes decrease as the FIR wavelength increases. These results
should be interpreted in terms of the increasing contribution of the
diffuse component with increasing FIR wavelength.

Using the ISOPHOT observations from Tuffs et al. (2002a,b) and
Stickel et al. (2000) in combination with UV and K band photometry,
Pierini \& M\"oller (2003) have tried to quantify the effects of
optical heating and disk opacity on the derivation of SFR from FIR
luminosities.  For this they investigated trends in the ratio of the
far-IR luminosity to the intrinsic UV luminosity, L$_{dust}$/L$_{UV}$,
with both disk opacity and disk mass (as measured by the intrinsic
K-band luminosity). Using a separate relation between disk opacity and
K band luminosity they were able re-express L$_{dust}$/L$_{UV}$ in
terms of a single variable, the galaxy mass
In this way they found evidence for the
relative importance of optical photons in heating dust to increase
with increasing galaxy mass.

\begin{figure} 
\includegraphics[width=\textwidth]{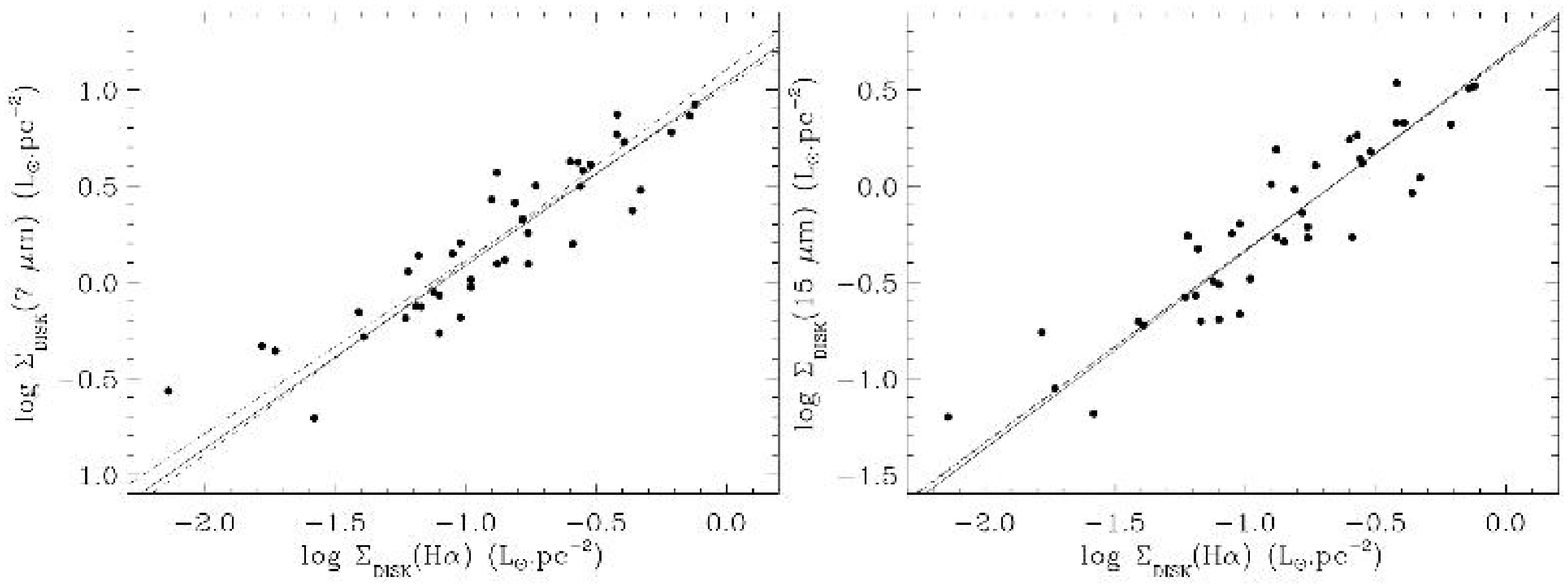}

\caption[]{The MIR-\hal\ correlation for the disks of spiral galaxies
(adapted from \opencite{rousfr}). Here only the disk fraction of the
MIR flux is used, and the \hal\ fluxes are uniformly corrected for
1\,magnitude of extinction. All fluxes are normalised by the galaxy
size to remove size-related biases. For both fluxes the correlation is
extremely linear. Lines on the graphs correspond to different adjustment
methods, but all give a slope very close to unity.}

\label{fig:f11}
\end{figure}

In the MIR, the correlation of global MIR and \hal\
luminosities is also non-linear \cite{rousfr}, but since the MIR SED
is so constant from galaxy to galaxy, an explanation along the lines
developed for the FIR correlation is not possible. However,
\inlinecite{rousfr} have shown that the non-linearity is due again
to the fact that the global MIR flux mixes together the contributions
from the central regions and the disk. Restricting their studies to
the disk emission, they showed that both the 6.75\,\mic\ and the
15\,\mic\ luminosities are linearly correlated with the
extinction-corrected \hal\ luminosities (see
Fig.~\ref{fig:f11}). Therefore, although the MIR emission from the
disk of spiral galaxies does not originate directly in the
star-forming regions, but rather from the PDRs around them, it is the
energy from the star-formation process that powers the MIR luminosity,
and Roussel et al. provide the conversion factors that allow the derivation of
the star-formation rate (SFR) from the MIR luminosities. In a
follow-up study, \inlinecite{for04} demonstrated that the 6.75\,\mic\
luminosity is still as good a tracer of star-formation in the central
regions of spiral galaxies as it was in their disks. The non-linearity
observed in the global 6.75-\hal\ diagram is attributed to the
difference in extinction between disk and central star-forming
regions. For the 15\,\mic\ luminosities, \inlinecite{for04} showed
that the star-formation activity observed in the central regions of
galaxies is such that it shifts the emission from very small grains
(usually emitting in the 20-60\,\mic\ range) toward shorter
wavelengths. Therefore, even when the \hal\ emission is corrected for
the higher central extinction, the correlation with the 15\,\mic\
emission is not linear as grains that make the IR emission
change their thermal regime.

\subsection{Quantitative interpretation of FIR SEDs}
\label{ssec:quant}

When considered in isolation, the FIR luminosity of normal galaxies is
a poor estimator of the SFR.
There are two reasons for
this.  Firstly, these systems are only partially opaque to the UV
light from young stars, exhibiting large variations in the escape
probability of UV photons between different galaxies. Secondly, the
optical luminosity from old stellar populations can be so large in
comparison with the UV luminosity of young stellar populations that a
significant fraction of the FIR luminosity can be powered by optical
photons, despite the higher probability of absorption of UV photons
compared to optical photons.

\begin{figure}[htb] 
\includegraphics[scale=0.55]{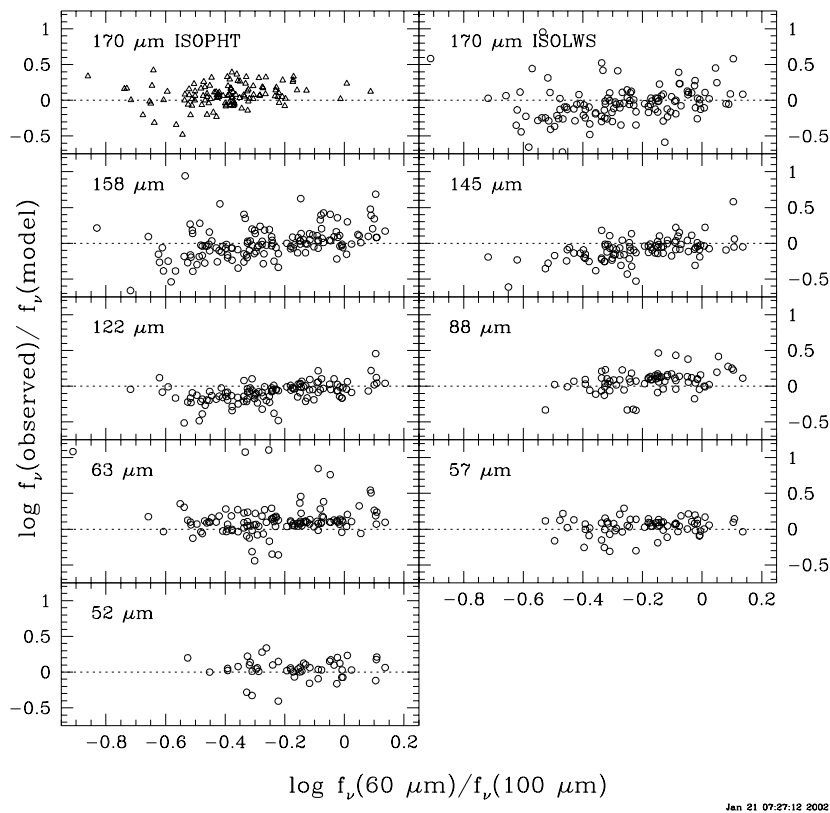}

\caption{Comparison of observed FIR continuum levels observed
by ISO with the model predictions of Dale \& Helou (2002). The circles
derive from the ISO LWS templates and the triangles represent
170\,${\mu}$m data from the ISOPHOT Serendipity Survey (Stickel et
al. 2000). }

\label{fig:f12}
\end{figure}

One step towards a quantitative interpretation of FIR SEDs was
achieved by the semi-empirical models of Dale et al. (2001) and Dale \&
Helou (2002). These authors have used ISO observations to develop a
family of templates to fit the variety of the observed forms of the IR
SEDs. This work assumes a power law distribution of dust masses with
local radiation field intensity to provide a wide range of dust
temperatures. It appears to indicate that the SEDs of star-forming
galaxies can be fitted by a one-parameter family of curves
(characterised by the 60/100\,${\mu}$m colour), determined essentially
by the exponent, $\alpha$, of the power law distribution of dust
masses with the radiation field intensity, where the radiation field
is assumed to have the colour of the local ISRF. This calibration
method has been extensively used to predict FIR flux densities
longwards of 100\,${\mu}$m for the galaxies not observed by ISOPHOT. A
comparison of the FIR continuum levels observed by ISO with
the Dale \& Helou (2002) model predictions is shown in
Fig.~\ref{fig:f12}.  However, a quantitative interpretation of dust
emission in terms of SFRs and star-formation histories
requires a combined analysis of the UV-optical/FIR-submm SEDs,
embracing a self-consistent model for the propagation of the photons.

\begin{figure}[htb] 
\includegraphics[scale=0.53]{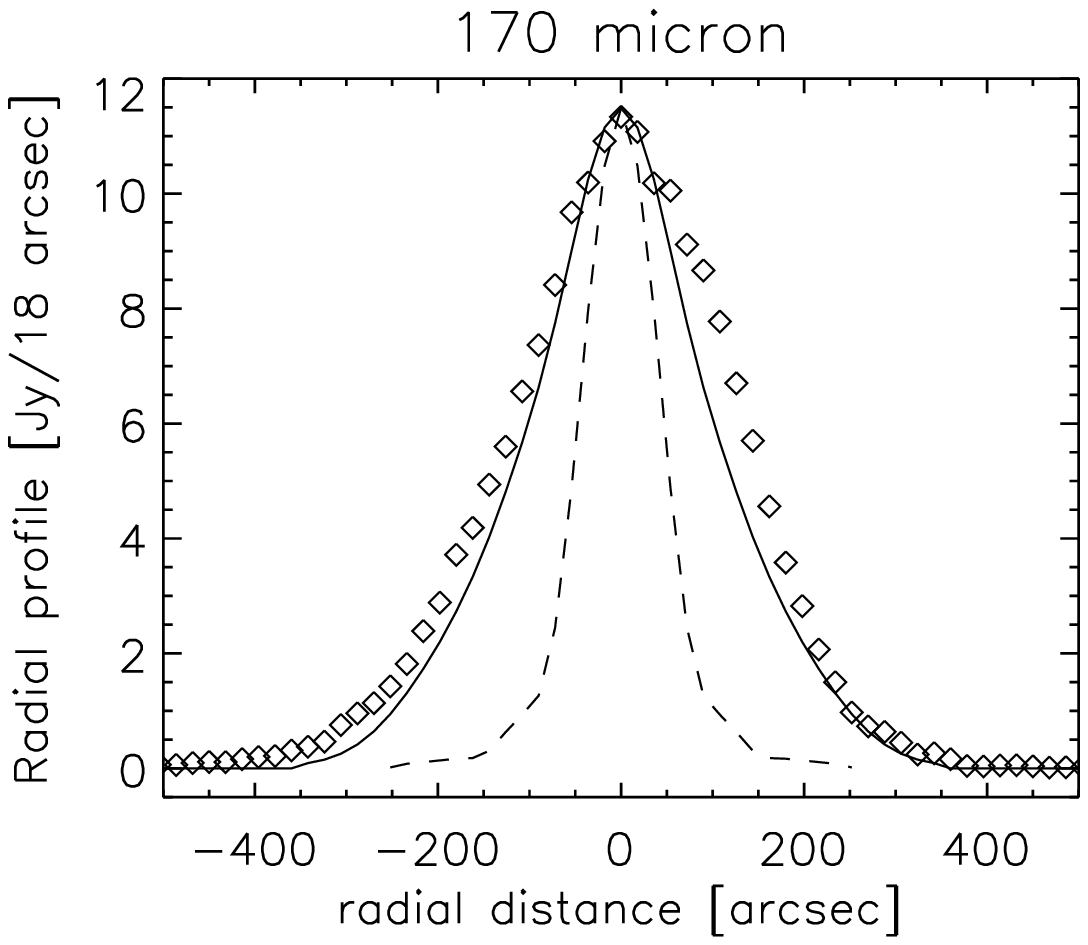}
\includegraphics[scale=0.40]{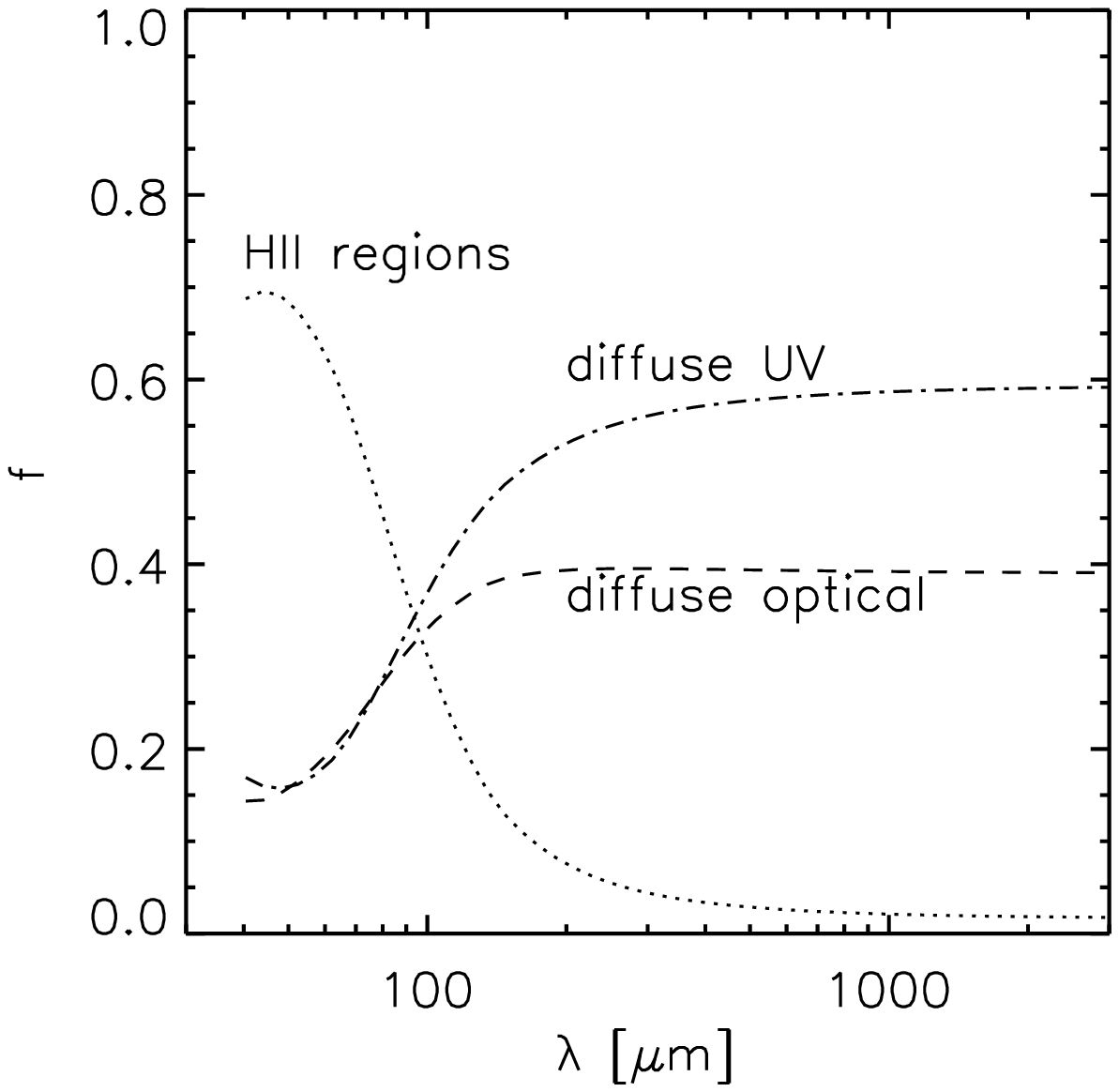}

\caption{Left: The radial profile of NGC~891 at 170\,${\mu}$m (Popescu et
al. 2004) produced by integrated the emission parallel to the minor
axis of the galaxy for each bin along the major axis. Solid line:
model prediction; diamonds: observed profile; dotted line: beam
profile.
Right: The fractional contribution of the three stellar components to the FIR
emission of NGC~891 (Popescu et al. 2000b).}

\label{fig:f13}
\end{figure}

There are a few such models in use which incorporate various
geometries for the stellar populations and dust (Silva et al. 1998,
Bianchi et al. 2000a, Charlot \& Fall 2000, Popescu et
al. (2000b)).  We will concentrate here on the model of Popescu et
al., since this is the only model which has been used to make direct
predictions for the spatial distribution of the FIR emission for
comparison with the ISO images. This model also ensures that the
geometry of the dust and stellar populations is consistent with
optical images. Full details are given by Popescu et al. (2000b),
Misiriotis et al. (2001), Popescu \& Tuffs
(2002b) and Tuffs et al. (2004).
In brief, the model includes solving
the radiative-transfer problem for a realistic distribution of
absorbers and emitters, considering realistic models for dust, taking
into account the grain-size distribution and stochastic heating of
small grains and the contribution of \hii\ regions.  The FIR-submm SED
is fully determined by just three parameters: the star-formation rate
$SFR$, the dust mass $M_{\rm dust}$ associated with the young stellar
population, and a factor $F$, defined as the fraction of non-ionising
UV photons which are locally absorbed in \hii\ regions around the
massive stars. A self-consistent theoretical approach to the calculation of the
$F$ factor is given by Dopita et al. (2005) in the context of modelling SEDs
of starburst galaxies.

Popescu et al. (2000b) illustrated their model with the example of the
edge-on galaxy NGC~891, which has been extensively observed at all
wavelengths (including the complete submm range; Dupac et al. 2003),
and also mapped with ISOPHOT at 170 and 200\,${\mu}$m (Popescu et
al. 2004).
A particularly stringent test of the model was to compare its
prediction for the radial profile of the diffuse dust emission
component near the peak of the FIR SED with the observed radial
profiles. This comparison (see Fig.~\ref{fig:f13}, left panel) done in
Popescu et al. (2004) showed a remarkable agreement.
The excess emission in the observed profile with respect to
the predicted one for the diffuse emission was explained in terms of two
localised sources.

At 170 and 200\,${\mu}$m the model for NGC~891 predicts that the bulk
of the FIR dust emission is from the diffuse component. The close
agreement between the data and the model predictions, both in
integrated flux densities, but especially in terms of the spatial
distribution, constitutes a strong evidence that the large-scale
distribution of stellar emissivity and dust predicted by the model is
in fact a good representation of NGC~891. In turn, this supports the
prediction of the model that the dust emission in NGC~891 is
predominantly powered by UV photons.

Depending on the FIR/submm wavelength, the UV-powered dust emission
arises in different proportions from within the localised component (\hii\
regions) and from the diffuse component (Fig.~\ref{fig:f13}; right
panel).  For example at 60\,${\mu}$m, 61$\%$ of the FIR emission in NGC~891 was
predicted to be powered by UV photons locally absorbed in star-forming
complexes, 19$\%$ by diffuse UV photons in the weak radiation fields
in the outer disk (where stochastic emission predominates), and 20$\%$
by diffuse optical photons in high energy densities in the inner part
of the disk and bulge. At 100\,${\mu}$m the prediction was that there
are approximately equal contributions from the diffuse UV, diffuse
optical and locally absorbed UV photons. At 170, 200\,${\mu}$m and
submm wavelengths, most of the dust emission in NGC~891 was predicted
to be powered by the diffuse UV photons. The analysis described above
does not support the preconception that the weakly heated cold dust
(including the dust emitting near the peak of the SED sampled by the
ISOPHOT measurements presented here) should be predominantly powered
by optical rather than UV photons. The reason is as follows: the
coldest grains are those which are in weaker radiation fields, either
in the outer optically thin regions of the disk, or because they are
shielded from radiation by optical depth effects. In the first
situation the absorption probabilities of photons are controlled by
the optical properties of the grains, so the UV photons will dominate
the heating. The second situation arises for dust associated with the
young stellar population, where the UV emissivity far exceeds the
optical emissivity.

\subsection{The gaseous ISM}
\label{ssec:gas}

\subsubsection{The atomic component}

Most ISO studies of the ISM in normal galaxies focussed on
observations with LWS of the [C~II] 158\,${\mu}$m fine structure
transition, which is the main channel for cooling of the diffuse
ISM.  Prior to ISO, knowledge of [C~II] emission from
external galaxies was limited to measurements of a total of 20
strongly star-forming galaxies sufficiently bright to be accessible to
the Kuiper Airborne Observatory (Crawford et al. 1985; Stacey et
al. 1991; Madden et al. 1993). The main conclusion of these early
investigations was that the [C~II] line emission arose from
PDRs, where the UV light from young stars
impinged on molecular clouds in star-forming regions.
The only exception to this was Madden et al's resolved study of NGC~6946,
who showed that the line could also be emitted from the diffuse
disk.

The principal contributions of ISO to the knowledge of the
neutral gaseous ISM have been twofold: Firstly,
knowledge of the [C~II] line emission has been extended to
quiescent spiral galaxies and dwarf galaxies. These targets are
difficult to access from beneath the atmosphere on account of
their low luminosities and brightnesses. Secondly, ISO has
tripled the number of
higher luminosity systems with [C~II] measurements, and has
also observed a whole range of other fine structure lines
in these systems. These lines have been used to probe in detail the
identity of the emitting regions and their physical parameters.

Fundamental to the studies of the quiescent spiral galaxies
were the measurements by Leech et al. (1999), who
detected 14 out of a sample of 19 Virgo cluster spirals (a subset of the
IVCDS sample - see Sect.~\ref{ssec:int}) in the [C~II] line using LWS.  These
galaxies are the faintest normal galaxies ever measured in the [C~II]
line, and can be regarded as quiescent in star-forming activity
(\hal\ EW is less than 10\,\AA\ for 8 galaxies). In a
series of papers Pierini \& collaborators (Pierini et al. 1999;
Pierini et al. 2001, \& \opencite{pie03a}) used these LWS
data, in combination with the existing KAO measurements of more
actively star-forming galaxies, to investigate the physical origin of
the line emission in normal galaxies as a function of star-formation
activity.  The main result was that the [C~II] emission of quiescent
systems predominantly arises from the diffuse ISM (mainly the cold
neutral medium; CNM), as evidenced from the fact that these systems
occupy a space in the L$_{[C~II]}$/L$_{CO}$ versus
L$_{[C~II]}$/L$_{FIR}$ diagram devoid of individual star-formation
regions or giant molecular clouds (Pierini et al. 1999).
An origin of the [C~II] emission
from the CNM had earlier been proposed by Lord et al. (1996) as
an alternative to an origin in localised PDRs, on
the basis of a full LWS grating spectrum of the spiral galaxy NGC~5713.

\begin{figure}[htb] 
\includegraphics[scale=0.40]{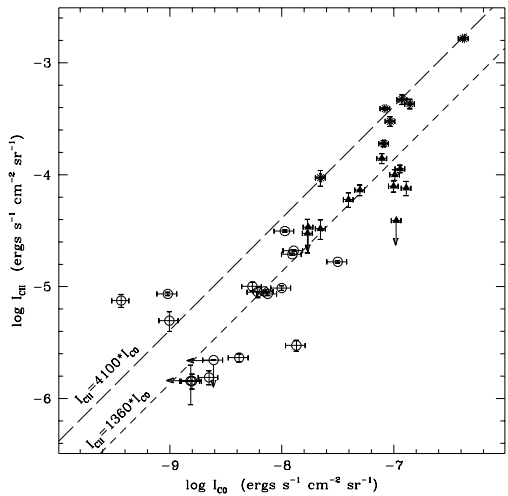}
\caption{The relationship between the observed central [CII] line
intensity, $I_{[CII]}$, and the central CO line intensity, $I_{CO}$
from Pierini et al. (2001). Asterisks and filled triangles denote
starburst galaxies and gas-rich galaxies of the KAO sample, while open
circles identify spiral galaxies of the ISO sample. The long- and
short-dashed lines show the average ratios of the two observables
obtained by Stacey et al. (1991) for the starburst galaxies and the
gas-rich galaxies, respectively.}
\label{fig:f14}
\end{figure}
The luminosity ratio of [C~II] line emission
to FIR dust emission, L$_{[C~II]}$/L$_{FIR}$, was found to be
in the range 0.1 to 0.8$\%$ for the quiescent spirals. This is
consistent with the basic physical interpretation,
originally established for the more active galaxies,
of a balance between gas cooling (mainly through the [C~II] line) and gas
heating through photoelectric heating from grains\footnote{The
exception is the
Virgo cluster galaxy NGC~4522, which was found by Pierini et
al. (1999) to have an abnormally high value for
L$_{[C~II]}$/L$_{FIR}$, possibly indicating mechanical heating of the
interstellar gas as the galaxy interacts with the intracluster
medium.} However, the range of a factor of almost an order of magnitude
in the observed values
of L$_{[C~II]}$/L$_{FIR}$ also suggests that the relation between the [C~II]
line emission and the SFR may be quite complex. This impression is
reinforced by the large scatter and non-linearities found by Boselli et al.
(2002) in the relations between the [C~II] luminosity and
SFRs derived from \hal\ measurements for a sample
principally composed of the Virgo galaxies measured by \inlinecite{lee99}
with the LWS.

One consequence of the physical association of the
[C~II] emission with the diffuse dusty ISM in quiescent galaxies
is that the observed L$_{[C~II]}$/L$_{FIR}$ ratio will be reduced
due to an optically- (rather than UV-) heated component of the FIR dust
emission. A model quantifying this effect was developed by
\inlinecite{pie03a} in which
L$_{[C~II]}$/L$_{FIR}$ depends on the fractional amount of the
non-ionising UV light in the interstellar radiation field in normal
galaxies. Overall, systematic variations in the L$_{[C~II]}$/L$_{FIR}$ ratio
for star-forming galaxies can be summarised as follows:
In progressing from low to high star-formation activities,
L$_{[C~II]}$/L$_{FIR}$ first increases in systems in which the
[C~II] emission is mainly from the diffuse medium, due to a decrease in
importance of optical photons in heating the diffuse dust. After reaching a
maximum, L$_{[C~II]}$/L$_{FIR}$ then decreases as the star-formation activity
is further increased to starburst levels, due to an increase in the
fraction of [C~II] emission arising from localised PDRs associated with
star-forming regions, coupled with the quenching of the [C~II] emission
through the decreased efficiency of photoelectric heating of the gas
in high radiation fields. Thus, the dominance of the [CII] emission from PDRs
turned out to be the asymptotic limit for high SFR in gas-rich galaxies.

The LWS observations of Virgo spirals also showed that the linear
relation between L$_{[C~II]}$ and L$_{CO}$ previously established for
starburst systems extends into the domain of the quiescent spirals,
though with an increased scatter (see Fig.~\ref{fig:f14}). The
tight relation between
L$_{[C~II]}$/L$_{CO}$ and the \hal\ equivalent width (EW) found by
Pierini et al. (1999) indicated that this scatter may be induced by different
strengths of the far-UV radiation field in galaxies. Alternatively,
as discussed by Smith \& Madden (1997) in
their study of five Virgo spirals, the fluctuations in
L$_{[C~II]}$/L$_{CO}$ from one galaxy to the next might
reflect changes in L$_{CO}$ from PDRs caused by variations in metallicity
(and dust abundance), as well as varying fractions of [C~II] emission
arising from PDRs and the cold neutral medium.
The measurements of the [C~II]/CO line ratios derived from ISO observations
have also thrown new light on the much debated relation between the
strength of the CO line
emission and the mass of molecular hydrogen in galaxies. Bergvall et al. (2000)
emphasise how the low metallicity, the intense radiation field and the low
column density in the dwarf starburst galaxy Haro 11 can explain the
extremely high observed [C~II]/CO flux ratio, indicating that CO may be a
poor indicator of the H$_2$ mass in such systems. Similarly, the
enhanced [C~II]/CO
ratios found in the Virgo spirals observed by Smith \& Madden (1997) were
attributed to the low metallicities in these galaxies, although, as also
pointed out by these authors, an alternative explanation could be that
radiation from diffuse HI may dominate the [C~II] emission.

Mapping observations of the [C~II] line using the LWS provide
a means to directly probe the relative amounts of
[C~II] arising from localised PDRs associated with star-forming regions
in the spiral arms and the diffuse disk, as well to investigate
the relation between the [C~II] emission and the neutral and molecular
components of the gas. They have also allowed the nuclear
emission to be separately studied. Contursi et al. (2002) mapped the two nearby
late-type galaxies NGC~1313 and NGC~6946
in the [C~II] line, finding that the diffuse HI disk contributes
$\le \sim $40~$\%$ and $\sim\,$30~$\%$ of the integrated [C~II] emission
in NGC~6946 and NGC~1313, respectively. CO(1-0) and [C~II] were also found
to be well correlated in the spiral arms in NGC~6946, but less well so
in NGC~1313. Stacey et al. (1999) mapped the barred spiral M83 in a variety
of fine structure lines in addition to the [C~II] line - the
[O~I] 63~\&\,145\,${\mu}$m lines, the [N~II] 122\,${\mu}$m line
and the [O~III] 88\,${\mu}$m line, and obtained a full grating
scan of the nucleus. At the nucleus, the line ratios indicate a strong
starburst headed by O9 stars. Substantial [N~II] emission from low density
HII regions was found in addition to the anticipated component from
PDRs. A further resolved galaxy studied with the LWS is M33, for which
Higdon et al. (2003) made
measurements of fine structure lines and dust continuum emission
towards the nucleus and six giant HII regions.
Overall, the picture presented by these investigations of resolved galaxies
is broadly in accordance with the theory that
the integrated [C~II] emission from spirals galaxies is comprised of
the sum of components from the diffuse disk and from localised PDRs
and diffuse ionised gas
associated with star-formation regions in the spiral arms.

Fundamental to the statistical investigations of gas in more active
galaxies are the full grating
spectra of 60 normal galaxies from
the ISO key-project sample of Helou et al. (1996), which were obtained by
Malhotra et al. (1997, 2001) with the LWS. One of the main results was
the discovery of a smooth but drastic
decline in L$_{[C~II]}$/L$_{FIR}$ as a function of increasing
star-formation activity (as traced by the L$_{60}$/L$_{100}$ IRAS colour
ratio or L$_{FIR}$/L$_{B}$) for the most luminous
third of the sample. This trend was
accompanied by a increase in the ratio of the luminosity of the
[O~I] 63\,${\mu}$m line to that of the [C~II] line. This is readily
explained in PDR models in terms of a decreased efficiency of photoelectric
heating of the gas as the intensity of the UV radiation field is increased,
coupled with an increased importance of the [O~I] cooling line as the gas
temperature is increased. Both Malhotra et al. (2001) and
Negishi et al. (2001) present a comprehensive analysis of the
ratios of the most prominent lines ([C~II] 158\,${\mu}$m;
[O~I] 63~\&\,145\,${\mu}$m; [N~II] 122\,${\mu}$m;
[O~III] 52~\&\,88\,${\mu}$m and [N~III] 57\,${\mu}$m) in terms of
the densities of gas and radiation fields in PDRs, as well as of
the fraction of emission that comes from the diffuse ionised medium (as traced
by [N~II] 122\,${\mu}$m). Malhotra et al (2001) found that the radiation
intensities $G_{0}$ increases with density $n$ as $G_{0} \propto n^{\alpha}$,
with $\alpha$ being $\sim1.4$. They interpret this result, together
with the high PDR temperatures and pressures  needed to fit the data, as
being consistent with the hypothesis that most of the line and continuum
luminosity of relatively active galaxies arises from the immediate proximity
of the star-forming regions.
By contrast, on the basis of a comparison of the [C~II] line with the
[N~II] 122\,${\mu}$m line, Negishi et al. (2001) argue that
a substantial amount (of order 50$\%$) of the [C~II] line emission
might arise not from PDRs but instead from low density diffuse ionised
gas. Negishi et al. (2001) also found a linear (rather than non-linear)
relation between $G_{0}$ and $n$, which would argue against a decreased
photoelectric heating efficiency being responsible for the decline
in  L$_{[C~II]}$/L$_{FIR}$ with star-formation activity. Instead,
Negishi et al. invoke either an increase in the collisional
de-excitation of the line with increasing density, or a decrease
in the diffuse ionised component with increasing star-formation
activity as being responsible for this effect.

An explanation for the decrease in L$_{[C~II]}$/L$_{FIR}$ in terms
of a decreased efficiency of photoelectric
heating of the gas (Malhotra et al. 2001) or due to a collisional
de-excitation of the [C~II] line (Negishi et al. 2001)
would predict that
the L$_{[C~II]}$ will not be a good probe of high redshift starburst galaxies.
However, an alternative (or complementary)
explanation for the decrease in L$_{[C~II]}$/L$_{FIR}$ with star-formation
activity, proposed by  Bergvall et al. (2000), is that the line
becomes self-absorbed in local universe starbursts with
high metallicity. This scenario would also explain why the metal poor
luminous blue compact dwarf galaxy Haro 11 was found by Bergvall et al.
to have a high ratio of L$_{[C~II]}$/L$_{FIR}$, despite the intense
radiation fields present.

\subsubsection{The molecular component}

The lowest energy transitions of the most abundant molecule in
the Universe, H$_{2}$, are located in the MIR, typically between 3 and
30\,\mic. These pure rotational lines potentially offer a new access
to the molecular component of the ISM, to be compared with the less
direct CO tracer. One caveat though is that the lines can be excited
by different mechanisms, typically shocks or UV-pumping, which render
their interpretation difficult. Furthermore, the emission is very
rapidly dominated by that from the warmest fraction of the gas, and
therefore this new tracer offers only a partial and complex access to the
molecular phase.

Most of the extragalactic H$_{2}$ detections were made with ISOSWS
(\opencite{tdg96}; \opencite{kl03})), since high spectral
resolution is necessary to isolate the very thin molecular lines
from the broader features that abound in the MIR regime (although
a tentative detection in NGC\,7714 made with ISOCAM is presented
by \opencite{oha00}). For sensitivity reasons, the galaxies that
were observed in these lines tended to be starburst or active
galaxies and are discussed elsewhere in this volume.

Nevertheless, the S(0) and S(1) transitions were observed in the
central region and in the disk of NGC\,6946 (\opencite{val96} and
\opencite{val99a}) and all along the disk of NGC\,891
\inlinecite{val99b}. In the central region of NGC\,6946, the molecular
component detected with ISOSWS is clearly the warm fraction (of order
5-10\%) of the molecular gas present in the region and is raised to
the observed temperature of 170\,K by the enhanced star-formation
activity in the center of the galaxy. H$_{2}$ is also detected further
out (4\,kpc) in the disk, with a cooler temperature, though still
relatively warm.

The observations of H$_{2}$ in the disk of NGC\,891 reach much larger
distances, extending about 10\,kpc on both sides of the
nucleus. \inlinecite{val99b} argue that the observed line ratios
indicate the presence of both warm (150-230\,K) molecular gas,
identical as that observed in NGC\,6946, and cool (80-90\,K) molecular
gas. This cooler gas needs to have a very large column density to
explain the observed surface brightness and the masses derived by
\inlinecite{val99b} are in fact so large that they would resolve the
missing mass problem in the optical disk of NGC\,891. This highly
controversial conclusion however rests on a rather uncertain basis
dealing with the actual spatial distribution of the H$_{2}$-emitting
clouds within the ISOSWS beam. Upcoming observations with the IRS on
Spitzer should remove the remaining uncertainties regarding the
presence of large amounts of cold H$_{2}$ in the disks of galaxies.

\section{Dwarf Galaxies}

\subsection{Cold dust surrounding dwarf galaxies}
\label{ssec:surround}

Gas-rich dwarf galaxies and in particular Blue Compact Dwarfs (BCDs)
were originally expected to have their FIR emission dominated by dust
heated locally in \hii\ regions. Temperatures of 30\,K or more were
anticipated. This was the a priori expectation in particular for the
BCDs, and became the standard interpretation for the IRAS results
obtained for these systems.  Hoffman et al. (1989), Helou et
al. (1988) and Melisse \& Israel (1994) each found that the
60/100\,${\mu}$m colours of BCDs were clearly warmer than those of
spirals.

The IVCD Survey (Tuffs et al. 2002a,b) changed that simple picture of
the FIR emission from dwarf galaxies.  The IVCDS included measurements
at 60, 100, and (for the first time) at 170\,${\mu}$m of 25 optically
selected gas-rich dwarf galaxies. The observations at 60 and
100\,${\mu}$m were consistent with the previous IRAS results on this
class of galaxies, though extending knowledge of these systems to
lower intrinsic luminosities. Unexpectedly however, high ratios of
170/100\,\mic\ luminosities were found in many of the surveyed
systems. Such long-wavelength excesses were found both in relatively
high-luminosity dwarfs, such as VCC655, as well as in fainter objects
near the limiting sensitivity of the survey. These observations imply
the presence of large amounts of cold dust.

As shown by Popescu et al. (2002), it seems unlikely that the cold
dust resides in the optically thick molecular component associated
with star-formation regions, since the implied dust masses
would be up to an order of magnitude greater than those typically
found in giant spirals.  Such large masses could not have been
produced through star-formation within the dwarfs over their
lifetime. One alternative possibility is that the dust
originates and still resides outside the optical extent of these
galaxies. In fact, some evidence for this was provided by the ISOPHOT
observations themselves, since they were made in the form of
scan maps, from which estimates of source sizes could be
determined. Even with the relatively coarse beam (1.6$^{\prime}$ FWHM at
170\,${\mu}$m), the extended nature of the sources could be clearly
seen in a few cases
for which the FWHM for the 170\,${\mu}$m emission exceeds the
optical diameters of the galaxies (to 25.5Bmag/arcsec$^2$) by factors
of between 1.5 and 3.5. It is interesting to note that two of these
galaxies (VCC~848 and VCC~81) have also been mapped in \hi\ (Hoffman et
al. 1996), revealing neutral hydrogen sizes comparable to the
170\,${\mu}$m extent. This raises the possibility that the
cold dust is embedded in the extended \hi\ gas, external to the optical
galaxy.  This would be analogous to the case of the edge-on spiral
NGC~891, where Popescu \& Tuffs (2003) discovered a cold-dust
counterpart to the extended \hi\ disk (see Sect.~\ref{sssec:firextent}). In
this context the main observational difference between the giant
spiral and the dwarfs may be that for the dwarfs the integrated
170\,${\mu}$m emission is dominated by the extended emission component
external to the main optical body of the galaxy, whereas for the giant
spirals the long-wavelength emission predominantly arises from within
the confines of the optical disk of the galaxy.

Apart from the SMC (which is discussed in Sect.~\ref{ssec:smc} and is
too extended for ISOPHOT to map beyond its optical
extent), only three dwarf galaxies were observed by ISOPHOT in the
field environment (all Serendipity Survey sources; Stickel et
al. 2000). All three sources have comparable flux densities at 100 and
170\,${\mu}$m. But the small statistics mean that it is still an open
question to which extent the cold dust emission associated with the
extended \hi\ component in dwarf galaxies is a cluster phenomenon or
not.

The existence of large quantities of dust surrounding gas-rich dwarf
galaxies may have important implications for our understanding of the
distant Universe.  According to the hierarchical galaxy formation
scenarios, gas-rich dwarf galaxies should prevail at the earliest
epochs. We would then expect these same galaxies to make a higher
contribution to the total FIR output in the early Universe, certainly
more than previously expected.

\subsection{Infrared emission from within dwarf galaxies}
\label{ssec:smc}

The distribution of cold dust within dwarf galaxies could be studied
in only one case, namely in the resolved (1.5$^{\prime}$ resolution)
170\,${\mu}$m ISOPHOT map of the Small Magellanic Cloud (Wilke et
al. 2003, Wilke et al. 2004).  The 170\,${\mu}$m ISO map of the SMC
reveals a wealth of structure, not only
consisting of filamentary FIR emitting regions, but also of numerous
(243 in total) bright sources which trace the bar along its major axis
as well as the bridge which connects the SMC to the LMC. Most of the
brighter sources have cold components, associated with molecular
clouds. The discrete sources were found to contribute $28\%$, $29\%$
and $36\%$ to the integrated flux densities at 60, 100 and
170\,${\mu}$m, respectively.
The SED was modelled by the superposition of 45\,K, 20.5\,K and 10\,K
blackbody components with emissivity index $\beta$=2. The average dust colour
temperature (averaged over all pixels of the 170/100 colour map) was found to
be $T_{\rm D}=20.3\,$K.

\inlinecite{bot04} have compared the FIR map of the SMC with
an \hi\ map of similar resolution. This reveals a good spatial
correlation of the two emissions in the diffuse regions of the maps
(regions that fall outside of the correlation are either hot star-forming
regions, or cold molecular clouds with no associated
\hi). Adding the IRAS data allows them to compute the FIR emissivity
per unit H atom. \inlinecite{bot04} found that this emissivity is lower
than in the Milky Way, and in fact it is even lower than the lower
metallicity ($Z_{\odot}/10$) of the SMC would imply, suggesting that
depletion mechanisms at work in the ISM have more than a linear
dependence on metallicity.

\begin{figure}[htb] 
\includegraphics[scale=0.5]{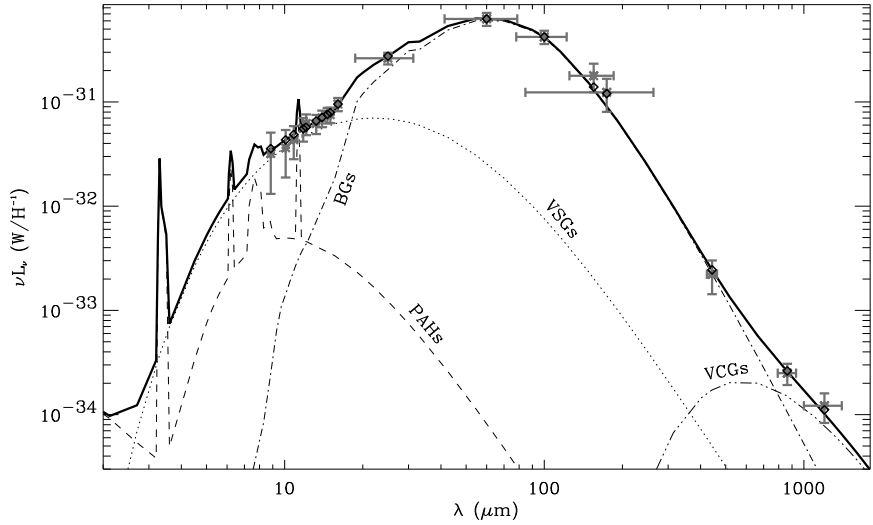}

\caption{NGC~1569 observations and modeled SED from Galliano et
al. (2003). The data are indicated by crosses: vertical bars are the
errors on the flux density values and the horizontal bars indicate the
widths of the broadbands. The lines show the predictions of the dust
model with its different components. Diamonds indicate the model
predictions integrated over the observational broadbands and
colour-corrected.}

\label{fig:f15}
\end{figure}

Although FIR emission from dwarf galaxies has been associated only with
gas-rich
dwarfs, in one particular case such emission has been detected in a dwarf
elliptical galaxy, as well. This is the case of NGC~205, one of the companions
of M~31, classified as a peculiar dE5. This galaxy shows signatures of recent
star-formation (Hodge 1973) and of extended HI emission (Young \& Lo 1997),
and was detected by IRAS (Rice et al. 1988, Knapp et al. 1989), with a SED
steeply rising between 60 and 100\,${\mu}$m. Based on
ISOPHOT observations, Haas (1998) showed that the FIR emission is
resolved and similar to that seen in HI. He also presented evidence for
a very cold dust component, of 10 K, coming from the center of the galaxy.

The properties of gas-rich dwarf galaxies were found to markedly differ
from those of  spiral galaxies at shorter IR wavelengths.
Thus, when comparing the MIR emission of five IBm galaxies to
their \hal\ or FIR properties, \inlinecite{hun01} discovered a
noticeable deficit of the 6.75\,\mic\ component in their
galaxies. This deficit is not observed at 15\,\mic, i.e. the
6.75/15\,\mic\ flux ratio is very different in IBm galaxies than in
spirals, which is interpreted as tracing both a deficit of PAH,
possibly destroyed by the radiation field, and an increase of the mean
temperature of the very small grains, again linked to the radiation
field (see also \opencite{oha00}). This interpretation of the
observations is confirmed by detailed spectroscopic measurements of
similar objects. In NGC\,5253 the MIR SED is dominated by a ``hot''
continuum from which the PAH features are absent \cite{cro99}. The
blue compact galaxy II\,Zw\,40 presents a similar spectrum, i.e. a
strongly rising continuum with no bands \cite{mad00}. In both cases
this is interpreted as being the result of very hard radiation fields
that both destroy the PAH and shift the continuum emission from very
small grains toward shorter wavelengths. These galaxies are
essentially small starburst regions and their IR spectrum thus
reflects the IR properties of starburst galaxies discussed
elsewhere in this volume. That we are indeed seeing a gradual
destruction of PAHs, and not an effect of metallicity that would
prevent the formation of these species is demonstrated by the fact
that we find galaxies of similar metallicity (NGC\,5253 and NGC\,1569
for instance at $Z_{\odot}/4-Z_{\odot}/5$) without or with PAH
emission (see \opencite{gal03} for NGC\,1569). In fact, even the SMC
at $Z_{\odot}/10$ has a MIR spectrum with pronounced PAH features
\cite{rea00}.


Many of the qualitative indications about grain and PAH abundance
indicated above have been quantitatively estimated by Galliano et al. (2003) in
their detailed modelling of the UV-optical/MIR-FIR-submm SED of the low
metallicity nearby dwarf galaxy NGC1569.
(Fig.~\ref{fig:f15}). This study is
noteworthy in that it constrains the grain size distribution through
the MIR-FIR ISOCAM and ISOPHOT observations and therefore gives
more specific information about grain properties in dwarf galaxies.
The results confirm the paucity of PAHs due to an enhanced destruction
in the
intense ambient UV radiation field, as well as an overabundance
(compared to Milky Way type dust) of small grains of size $\sim
3$\,nm, possibly indicative of a redistribution of grain sizes through
the effect of shocks.

\section{Early-type galaxies}
\label{sec:ellipt}

Here we use the term ``early-type'' to embrace both elliptical and
lenticular galaxies.
Their IR emission, when
detected, has been interpreted very differently from the IR emission from
spiral galaxies. Indeed, early-type galaxies belong to a subset of
the Hubble sequence where star-formation activity is at its minimum,
or has stopped, and the ISM is almost exhausted. Given what we have
seen above, one does not expect these objects to be IR
bright. This is indeed the case, judging from the small fraction of
early-type objects
in the IRAS catalogues (see
e.g. \opencite{jur87}), and the large uncertainties associated with
their fluxes \cite{bre98}. For those few galaxies where IR
emission is detected, it is commonly attributed either to the
Rayleigh-Jeans tail of the stellar emission, or to dust in mass-losing
stars. These interpretations however rest on the 4 broad-band low
spatial resolution data of IRAS. ISO, with its much improved
sensitivity to both spatial and spectral details, has 
shone a sharper
light on the subject.

The FIR observations of early-type galaxies selected from the ISO
archive have been analysed by Temi et al. (2004). They found that
the FIR SED requires emission from dust with at least two
different temperatures, with mean temperatures for the warm and
cold components of 43 and 20\,K, respectively. This result is
quite similar to the results obtained for quiescent spiral
galaxies, in particular to the result obtained by Popescu et al.
(2002) for the Virgo Cluster spiral galaxies. Since ellipticals
are known to be ISM-poor and have reduced star-formation activity,
the similarity between their FIR properties and those of spirals
is surprising. One explanation could be the fact that the galaxies
included in the ISO archive are not representative of typical
ellipticals. As pointed out by Temi et al., many galaxies selected
for ISO observations were chosen because they were known to have
large IRAS fluxes or large masses of cold gas, and in this sense
they have properties closer to those of spirals. For the same
reason Temi et al. considered that these IR-selected galaxies are
likely to have experienced unusual dust-rich mergers, and that the
dust in these galaxies could have an external origin. The lack of
a correlation found between the FIR luminosity (or dust mass) and
the luminosity of the B band emission was taken as evidence for
this scenario. Another scenario proposed is one in which most of
the FIR is emitted by central dust clouds, where the clouds could
be disturbed at irregular intervals by low-level AGN activity in
the galactic cores, creating the stochastic variation in the FIR
luminosity. It is difficult to distinguish between these two
scenarios, also because the sensitivity of ISO was not sufficient
to detect FIR emission from many optically luminous elliptical
galaxies. In order to distinguish between an external and an
internal origin for the dust, deeper FIR observations of an
optically selected sample of elliptical galaxies are needed. It is
even possible that such surveys may occasionally detect bright FIR
emission from galaxies traditionally classified as ellipticals,
but which in fact have hidden starburst activity in their central
regions. This is the case for the IRAS source F15080+7259, which
was detected in the ISOPHOT serendipity survey, and was shown by
Krause et al. (2003) to harbour large quantities of gas and cold
dust.

Another elliptical detected by ISOPHOT is the Virgo Cluster galaxy M~86
(NGC~4406). The deep FIR imaging data obtained with ISOPHOT at 60, 90, 150 and
180\,\mic\ revealed a complex FIR morphology for this galaxy
(Stickel et al. 2003). It was found that the FIR emission originates from
both the centre of M~86 and from the optically discovered dust streamers, and
has a dust temperature of ~18\,K. Again, this elliptical is not representative
for this class of objects, since
it was considered either to be experiencing ram-pressure stripping or to
be gravitationally interacting with its neighbouring galaxies. In particular
the ISOPHOT
observations were interpreted by Stickel et al. as being consistent with M~86
having a gravitational interaction with its nearest spiral galaxy NGC~4402. In
this case the detected dust would again have (at least in part) an external
origin.
The importance of gravitational interaction as a driver for FIR emission in
early-type galaxies was demonstrated by Domingue et al. (2003), who showed
these systems to have a 50$\%$ detection rate with ISOPHOT when they are
paired with spirals.

In the MIR ISOCAM was used to observe both elliptical and lenticular galaxies,
though the objects selected differ from those observed by ISOPHOT.
An unbiased sample of 16
IRAS-detected lenticular galaxies was
observed using low resolution MIR spectroscopy by \inlinecite{sau04}. The vast
majority of the detected galaxies present MIR SEDs extremely similar
to those of later-type spiral galaxies (i.e. PAH bands). Thus, from
their MIR spectral properties, lenticular galaxies appear to
belong to the spiral galaxy group, although with smaller luminosities.

Elliptical galaxies have drawn much more interest from ISOCAM observers,
even though they are ISM-poorer than the lenticulars.
The
main result from these studies is that elliptical galaxies are
really a mixed bag, as far as their MIR properties are
concerned. \inlinecite{qui99} mapped with ISOCAM a number of E+A
galaxies\footnote{Early-type galaxies showing evidence of a
post-starburst population.} in the Coma cluster, but detected only
those which showed emission lines characteristic of an on-going star-formation
 episode. This
might be thought to imply that star-formation is
required for ellipticals to emit in the MIR, but
in fact more likely reflects the sensitivity reached
by \citeauthor{qui99}. With deep ISOCAM
observations of closer ``normal'' elliptical galaxies,
\inlinecite{ath02} showed that their MIR emission is dominated by the
emission from K and M stars, with the addition of a spectral
feature near 10\,\mic\ that they attribute to silicates in the
circumstellar envelopes of AGB stars. \inlinecite{xil04} brought
balance to the star/ISM debate for the origin of MIR emission in
early-type galaxies by showing that the complete range of phenomena
occur in this class of galaxies. Systems vary from those that are
dominated up to
$\sim$20\,\mic\ by the emission from their old stellar populations to
those in which a small star-forming episode, possibly fueled by
recent accretion, dominates the MIR luminosities. In between one also
finds galaxies where only the long-wavelength part of the MIR spectrum
deviates from the extrapolation of the stellar light, which indicates
the presence of small amounts of diffuse dust much hotter than what is
commonly found in spirals. This is most likely to be due to the
higher
interstellar radiation field generated by the denser stellar
populations.

Although ISO has had some success in detecting and measuring the
IR SED from early-type galaxies, a critical look at the observations
done reveals that we still do not know the IR properties of
typical elliptical galaxies. To achieve this, systematic deep observations
of optically selected samples are required.

\section{Conclusions and Outlook}

ISO has not only advanced the knowledge of IR properties of normal
galaxies but has also made unexpected discoveries.
The bulk of the emission from dust has been measured,
revealing cold dust in copious quantities. This dust is present in all types
of normal galaxies and is predominantly distributed in a diffuse disk with an
intrinsic scalelength exceeding that of the stars. Cold
dust has been found beyond the optical regions of isolated
galaxies, associated with the extended HI disks of spiral galaxies or with the
HI envelopes of dwarf galaxies. The fraction of the
bolometric luminosity radiated by dust has been measured for the first time.
Realistic
geometries for stars and dust have been derived from ISO imaging
observations, enabling the contribution of the various stellar populations to
the dust heating to be accurately derived. The NIR/MIR SEDs were shown to be
remarkably similar for normal spiral galaxies and to originate mostly in the
PDRs surrounding star-forming regions.  A diffuse component of the
MIR emission was found, though puzzlingly with a much smaller scalelength
than its FIR counterpart. The importance of the central regions in shaping up
the intensity and the colour of the MIR global emission was revealed thanks to
 the high spatial resolution offered by ISO.  A new
component of interstellar dust emission consisting of a ``hot'' NIR continuum
emission was discovered. The relative
contribution of photospheric and very small grains/ PAH emission has been
established in the NIR/MIR spectral region. Spectroscopic observations
have revealed that the main
interstellar cooling line, [CII], is predominantly carried by the diffuse
cold neutral medium in normal galaxies, with emission from localised PDRs
only dominating for galaxies having high star-formation activity.

This enormous advancement in the understanding of normal galaxies in the
nearby Universe has laid the foundation for more detailed
investigations with Spitzer and Herschel.
A clear priority is to fill the gap left by ISO in
knowledge of the SEDs of normal galaxies between 20-60\,\mic. Another priority
is to increase the number of galaxies with detailed imaging information and to
provide better statistics on carefully selected samples, especially those
selected in the optical/NIR bands. Ultimately, the improved sensitivity of
the new
infrared space observatories will allow knowledge of the dust emission from
normal galaxies to be extended beyond the nearby universe.

\acknowledgements{
The authors would like to take this opportunity to thank all the individuals
that helped make the ISO mission a success. M. Sauvage acknowledges the
Max Planck Institut f\"ur Kernphysik for its support during the final
adjustments of the manuscript. R.J. Tuffs and C.C. Popescu would also like to
thank Heinrich J. V\"olk for enlightening discussions.}

\end{article}
\end{document}